\documentclass[twocolumn,amsmath,amssymb]{revtex4}
\pdfoutput=1
\usepackage{multirow}
\usepackage{hyperref}
\usepackage{graphicx}
\usepackage{dcolumn}
\usepackage{bm}
\usepackage{subfigure}
\newcommand{\be}{\begin{equation}}
\newcommand{\ee}{\end{equation}}
\newcommand{\bea}{\begin{eqnarray}}
\newcommand{\eea}{\end{eqnarray}}
\newcommand{\bw}{\begin{widetext}}
\newcommand{\ew}{\end{widetext}}

\begin{document}

\title{A Study of Hadron Deformation in Lattice QCD}

\author{Constantia Alexandrou}
\affiliation{
  Department of Physics, University of Cyprus, CY-1678, Cyprus 
}

\author{Giannis Koutsou}
\affiliation{
  Department of Physics, University of Cyprus, CY-1678, Cyprus 
}

\date{\today}

\begin{abstract}
  We develop the formalism for the evaluation of density-density correlators
  in lattice QCD that includes techniques for the computation
  of the all-to-all propagators involved. A novel technique in this context
  is the implementation of the one-end trick
  in the meson sector.
  Density-density correlators
  provide a gauge invariant definition for the 
  hadron wave function and yield information on hadron deformation.
  We evaluate density-density correlators using two degenerate flavors of
  dynamical Wilson fermions for
  the pion, the rho-meson, the nucleon and the $\Delta$. Using the
  one-end trick we obtain results that clearly
  show deformation of the rho-meson.
\end{abstract}

\maketitle

\section{Introduction}
Deformation in nuclei~\cite{Bohr, Lee:1975zz} and atoms~\cite{Berry,Ceraulo:1991} 
is an important phenomenon that has been 
extensively studied. In this work we address the question of whether deformation
also arises in low-lying hadrons using the fundamental theory of the strong 
interactions, Quantum Chromodynamics defined on the lattice. In order to
be able to answer this question we
develop techniques for the exact evaluation of four-point correlators. These methods are
also needed in a range of other applications in lattice QCD. 

In this work we study the shape of the pion, the rho-meson,
the nucleon (N) and the $\Delta$.
The pion being a spin-0 particle is expected to have no deformation
and it therefore provides a check for our methodology. For particles 
with spin larger than $1/2$, the
one-body quadrupole operator provides a convenient characterization of 
deformation.
The spin 
1/2 nucleon cannot have a 
spectroscopic quadrupole moment but can still have an intrinsic deformation.
The experiment of choice to reveal 
the presence of deformation in the nucleon and its excited state the $\Delta$ is measuring the
N to $\Delta$ transition amplitude. Significant effort has been 
devoted to photo- and electro-production 
experiments on the nucleon at major experimental facilities~\cite{Blanpied:1996yh,MAMI2:2001,Mertz:1999hp,Joo:2001tw}.
These experiments measure to high accuracy 
the ratios of the electric (E2) and Coulomb (C2) quadrupole amplitudes to 
the magnetic dipole (M1) amplitude.
If both the nucleon and the $\Delta$
are spherical, then E2 and C2 are
expected to be zero. 
There is mounting experimental evidence over a range of momentum transfers
that E2 and C2 are non-zero~\cite{Papanicolas:2003,Papanicolas:2006athens}.
These ratios have been recently shown to be non-zero in lattice QCD~\cite{Alexandrou:2007dt} pointing to deformation in the nucleon or/and $\Delta$.

A different approach that sheds light on deformation is to use density-density correlators to directly probe the hadron wave function~\cite{Alexandrou:2003ab,Alexandrou:2006athens}.
Density-density
correlators~\cite{Burkardt:1994pw,Gupta:1992rr,Alexandrou:2002nn,Alexandrou:2002st,Alexandrou:2002gt,Alexandrou:2003qt,Alexandrou:2003me,Alexandrou:2004ws,Alexandrou:2005dw,Alexandrou:2006pd,Alexandrou:2007pn}
provide a gauge invariant definition of the hadron wave function.
In a previous study~\cite{Alexandrou:2002nn}
the density-density correlators were evaluated approximately.
This was due to the fact that the all-to-all propagators
needed for their exact evaluation were not calculated.
Furthermore they were computed for pion masses larger than 600~MeV
and on lattices with a spatial volume of about 1.5~fm.

In this work we provide an exact evaluation of the four-point
functions involved in the computation of the density-density correlators. 
The all-to-all propagators needed for the exact evaluation are calculated
using stochastic techniques combined with dilution. 
In addition, we apply in the meson-sector for the first time in this context,
the so-called one-end trick originally devised to evaluate 
the pion zero momentum
two-point function~\cite{McNeile:2006bz}. 
In the two-point function, the one-end trick amounts to a clever summation of the spatial coordinates 
not only of the sink as routinely done but also of the source and therefore
all-to-all propagators are involved. Implementation of this
trick in the evaluation of the meson density-density correlators leads to a significant
reduction of the statistical errors~\cite{Alexandrou:2007pn}.
This trick, in its present formulation, can only be applied to meson density-density correlators.
In baryons, 
the density insertions are on only two of the three quarks which gives
rise to an odd number of quark propagators that cannot be grouped in pairs for the summation
to work.

An alternative method applicable to both mesons and
baryons is to combine stochastic evaluation of one all-to-all
propagator with a sequential inversion to sum over the other
spatial coordinate. This method, apart from the requirement of fixing the final hadronic state,
needing new sequential inversions for each of the nucleon and $\Delta$ states,
has been shown to
yield results with similar errors as using two sets of stochastic inversions~\cite{Alexandrou:2005dw} 
We therefore do not consider it here.

Further improvements as compared to the previous study of density-density 
correlators~\cite{Alexandrou:2002nn} is that we use a spatial lattice of $24^3$ 
as compared to $16^3$ used previously
and dynamical Wilson fermions corresponding to smaller pion masses, the lowest being 380 MeV. 

This paper is organized as follows: In Section II we define the density-density correlators,
in Section III
we explain the stochastic techniques used for the evaluation of the all-to-all propagators, 
in Section IV we
give the interpolating fields and parameters of the simulations and
in Section V we describe our results on the density-density correlators for the
pion, the rho-meson, the nucleon and the $\Delta$ and show how to correct for finite
spatial volume effects. Finally in Section VI we summarize and give our conclusions.

\section{Density - density Correlators}
Throughout this work we consider the equal-time density-density correlators defined by:
\begin{widetext}
  {\everymath{\displaystyle}
    \bea
    \label{Eq:DD_lattice} 
    \tilde{C}_{H}(\vec{x}_2,t_1)&=&\int d^3x_1 \langle H |j_0^u(\vec{x}_2+\vec{x}_1,t_1) j_0^d(\vec{x}_1,t_1)|H\rangle\nonumber \\
    &= &\int d^3x_1 \int d^3 x 
    \langle\Omega|J_H(\vec{x},t)j_0^u(\vec{x}_2+\vec{x}_1,t_1) 
    j_0^d(\vec{x}_1,t_1)J^\dagger_H(\vec{x}_0,t_0)|\Omega\rangle
    \eea
  }
  \noindent where $j_0^q$ is the normal ordered density operator $:\bar{q}\gamma_0q:$ and
  $J_H$ is an interpolating field with the quantum numbers of the lowest lying hadron $H$.
  The two integrals in Eq.~(\ref{Eq:DD_lattice}) ensure that the state
  is projected to zero momentum; one integral sets the momentum of the sink equal to that of the 
  source while the other sets both to zero. This can be shown explicitly by inserting three complete 
  sets of states in Eq.~(\ref{Eq:DD_lattice}):
  {\everymath{\displaystyle}
    \be
    \tilde{C}_{H}(\vec{x}_2,t_1)= \sum_{\vec{p},n,n_i,n_f} \langle\Omega|J_H|n_f,\vec{0}\rangle\>\>\frac{e^{-E_{n_f}(\vec{0})(t-t_1)}}{2E_{n_f}(\vec{0})}\>\>
    \langle n_f,\vec{0}|j_0^u|n,\vec{p}\rangle \>\>\frac{e^{i\vec{p}\cdot\vec{x}_2}}{2E_n(\vec{p})}\langle n,\vec{p}|j_0^d|n_i,\vec{0}\rangle \>\>
    \frac{e^{-E_{n_i}(\vec{0})(t_1-t_0)}}{2E_{n_i}(\vec{0})}\>\>\langle n_i,\vec{0}|J^\dagger_H|\Omega\rangle.
    \ee
  }
  In the large $t_1-t_0$ and $t-t_1$ limit we have:
  \bea
  C_H(\vec{x}_2)&=&{\rm Lim}_{(t-t_1)\rightarrow \infty;(t_1-t_0)\rightarrow \infty} \tilde{C}_H(\vec{x}_2,t_1)
  \nonumber \\
  &= & \sum_{\vec{p},n} |\langle\Omega|J_H|H\rangle|^2 \>\>\frac{e^{-m_H(t-t_0)}}{4m_H^2} \>\>
  \langle H|j_0^u|n,\vec{p}\rangle \>\>
  \frac{e^{i\vec{p}\cdot\vec{x}_2}}{2E_n(\vec{p})}\>\> \langle n,\vec{p}|j_0^d|H\rangle.
  \eea
\end{widetext}
If we divide by the zero momentum hadron two-point function $G_H(\vec{0},t-t_0)$ then the exponential 
dependence on $t-t_0$ and overlaps cancel and we obtain the expectation value of the two density insertions,
$\langle H|j_0^u(\vec{x}_2)j_0^d|H\rangle$.
In the non-relativistic limit, this expectation value gives
the charge distribution of the hadron. It can be written in terms of the non-relativistic
form factors~\cite{Burkardt:1994pw}
\be
\langle H|j_0^u(\vec{x}_2)j_0^d|H\rangle=\sum_{\vec{p},n} F^u_{Hn}(\vec{p})\>\>\frac{e^{i\vec{p}\cdot\vec{x}_2}}{2E_n(\vec{p})} \>\>F^d_{nH}(-\vec{p})
\label{matrix element}
\ee
where
\be
F^u_{Hn}(\vec{p})=\langle H|j_0^u|n,\vec{p}\rangle \quad.
\label{form factor}
\ee
\begin{figure}[!ht]
  \includegraphics[width = 0.7\linewidth]{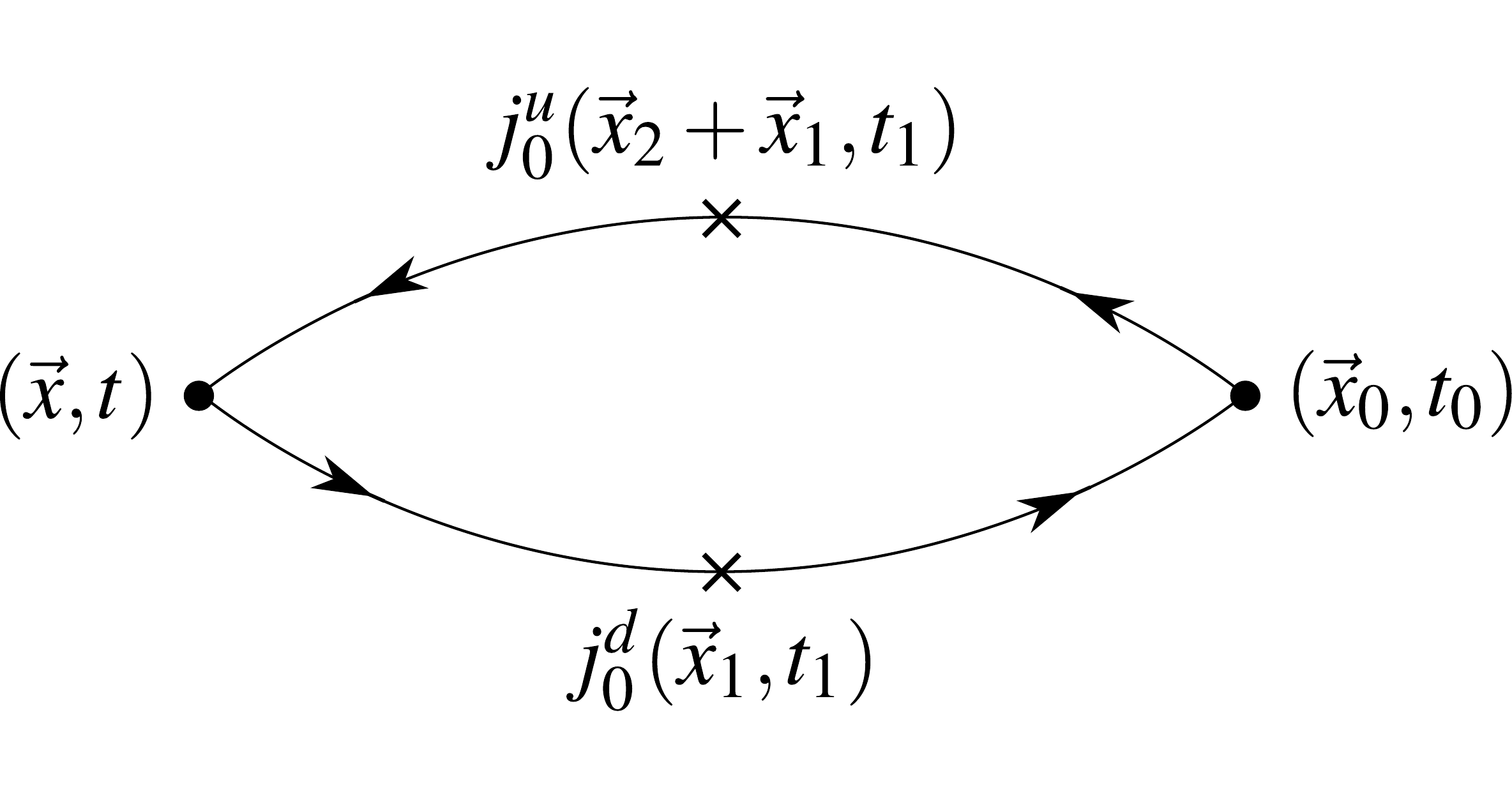}
  \includegraphics[width = 0.7\linewidth]{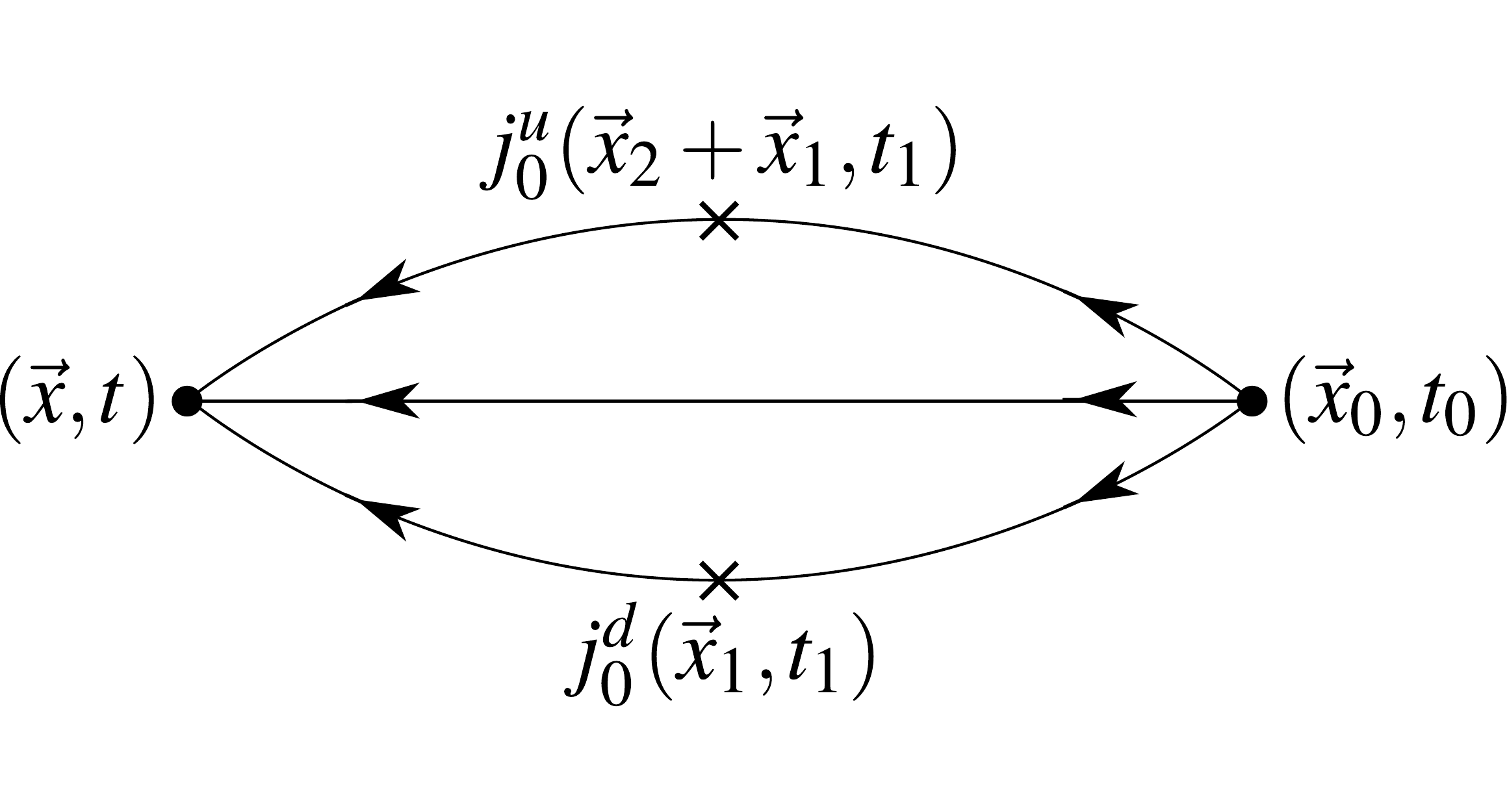}
  \caption{Equal-time density-density correlators for mesons (upper diagram) and for baryons (lower diagram). }
  \label{Fig:DDCorr_Diagram}
\end{figure}

\begin{figure}[!ht]
  \begin{minipage}{0.45\linewidth}
    \includegraphics[width = 0.9\linewidth]{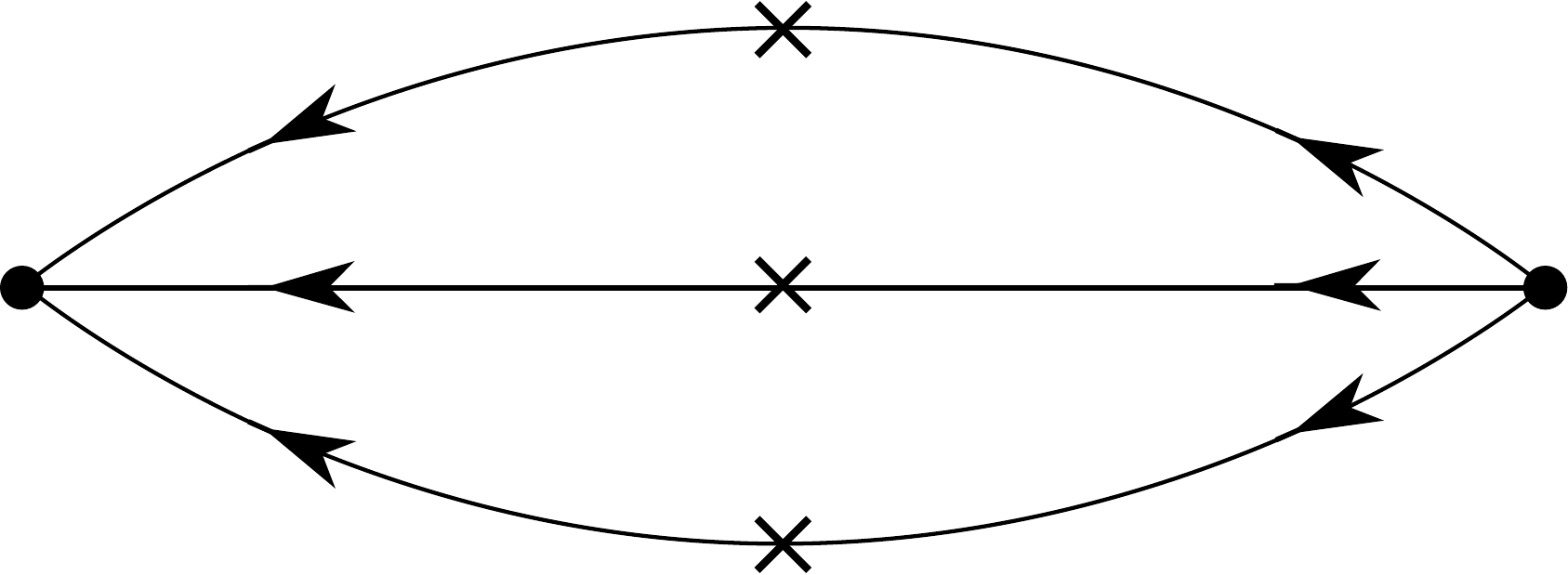}
  \end{minipage}
  \begin{minipage}{0.45\linewidth}
    \hfill
    \includegraphics[width = 0.9\linewidth]{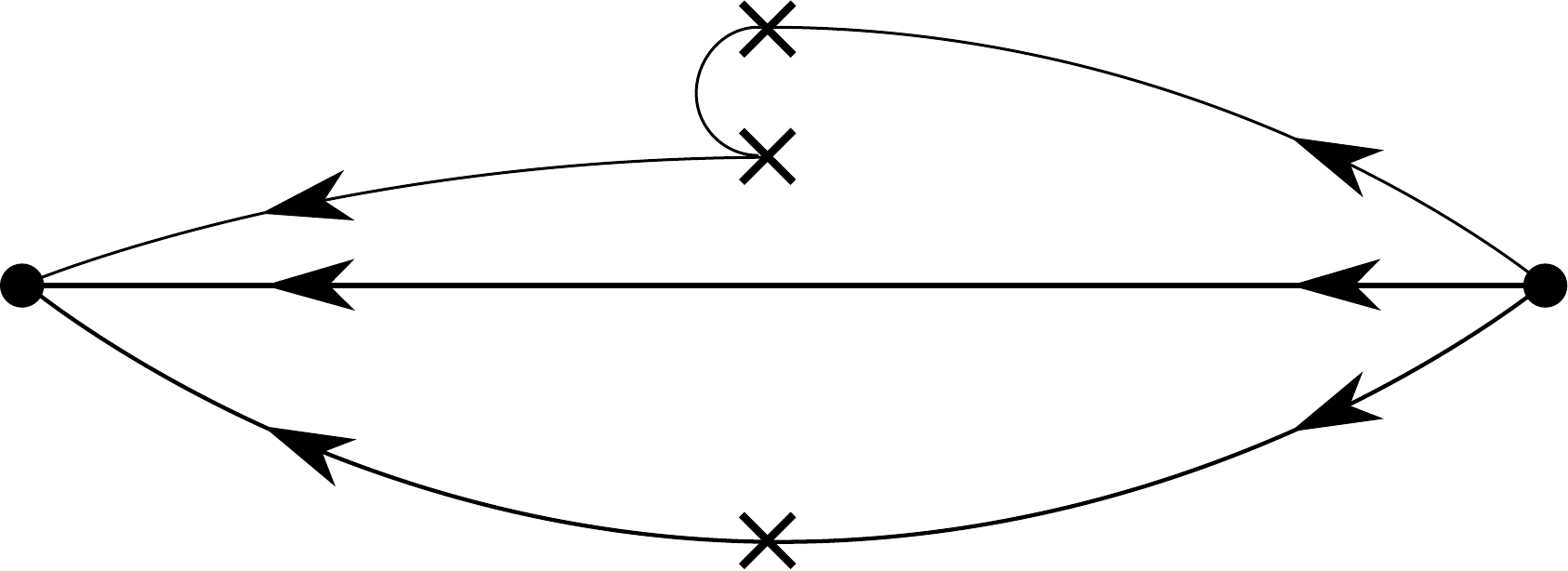}
  \end{minipage}
  \caption{The three density correlator for baryons.}
  \label{Fig:DDDCorr_Diagram}
\end{figure}

The connected diagrams of the density-density correlators for mesons and baryons are shown 
in Fig.~\ref{Fig:DDCorr_Diagram}.
We note here that the diagram depicted in Fig.~\ref{Fig:DDCorr_Diagram} for baryons yields 
a correlator that depends only on one relative distance instead of two. 
To obtain, in the non-relativistic limit, the charge distribution that depends on the two relative
distances one must calculate the three-density correlator.
This requires the evaluation of two types of five-point functions shown in Fig.~\ref{Fig:DDDCorr_Diagram}.
In Ref.~\cite{Alexandrou:2002nn} the three-density correlator or five-point function was
evaluated approximately for one of the diagrams shown in Fig.~\ref{Fig:DDDCorr_Diagram}
for which each quark line has only one density insertion. It was shown 
that integrating over one relative distance one obtains results that are consistent with the 
corresponding two-density correlator. 
For the work presented here we therefore only consider correlators with two-density insertions, which
give the distribution of one quark relative to the other irrespective of the position of the third. In
other words, in the non-relativistic limit, it corresponds to 
the one-body 
charge distribution.

What makes four-point functions 
harder to evaluate than three-point functions is the fact that we need to compute all-to-all
propagators. Sequential inversions used in the evaluation of three-point functions can not be used
here. The reason is that we are interested in obtaining the dependence
in terms of a relative distance and therefore the spatial positions where
the density operators are inserted involve the relative distance and
can not be summed independently. Therefore the bulk of this work deals with the
evaluation of the all-to-all propagators to sufficient accuracy.

\section{Stochastic Techniques}
The technically challenging aspect of the calculation of the density-density 
correlators is the fact that the summation over sink and insertion coordinates requires knowledge of all-to-all 
propagators. A previous study has been carried out in the quenched approximation
and using two dynamical degenerate Wilson fermions in which no 
summation was performed over the sink coordinates~\cite{Alexandrou:2002nn}. This eliminated the need of calculating 
all-to-all propagators at the cost of
not explicitly projecting to zero momentum states, which instead were 
only obtained via the large Euclidean time
suppression of higher momenta. In this work we 
use stochastic techniques to estimate the all-to-all propagators~\cite{Michael:1998sg,Collins:2007mh} enabling us to 
sum over the sink coordinate and thus explicitly project to zero momentum initial and final states.

In order to evaluate the all-to-all propagator one begins by defining an ensemble of $N_r$ 
noise vectors $\xi^a_\mu(\vec{x},t)_r$ obeying to order {$ \left(\frac{1}{\sqrt{N_r}}\right)$ 
  \bea
  \langle\xi_\mu^a(\vec{x},t)\rangle_r &=& 0\qquad\rm{and}\nonumber\\
  \langle\xi^a_\mu(\vec{x},t) \xi^{\dagger a^\prime}_{\mu^\prime}(\vec{x}^\prime,t^\prime)\rangle_r& =& \delta(\vec{x}-\vec{x}^\prime) \delta(t-t^\prime) \delta_{\mu \mu^\prime}\delta_{a a^\prime} 
  \label{Eq:Noise_Condition}
  \eea
  where $\mu$ and $a$ are spinor and color indices respectively and $r$ enumerates the vector in the stochastic 
  ensemble. In particular, we use Z(2) noise where $\xi_\mu^a(\vec{x},t)\in \{1.,i,-1,-i\}$ with 
  equal probability. By solving the Dirac equation with each of these $N_r$ noise vectors as the source, one 
  obtains an ensemble of solution vectors: 
  \be
  \phi^a_\mu(x)_r = \sum_y G^{ab}_{\mu\nu}(x;y) \xi^b_\nu(y)_r
  \label{Eq:Def_Sol}
  \ee
  where $\phi$ is a solution vector and $G$ is the inverse of the Dirac operator. If we now take the 
  average over the product between solution and noise vectors over the stochastic ensemble, we obtain an
  estimate of
  the all-to-all propagator:
  \bea
  \langle\phi^a_\mu(x)\xi^{\dagger b}_\nu(y)\rangle_r &=& \sum_{z} G^{a c}_{\mu \kappa}(x;z)\langle\xi^c_\kappa(z)\xi^{b \dagger}_\nu(y)\rangle_r\nonumber\\
  &=& \sum_{z} G^{a c}_{\mu \kappa}(x;z)\delta(z-y) \delta_{\kappa \nu}\delta_{c b}\nonumber\\
  &=& G^{a b}_{\mu \nu}(x;y).
  \eea
  
  A well known technique used to suppress stochastic noise is dilution \cite{Foley:2005ac}.
  Within this technique, one distributes the elements of a noise vector over certain color, spin and 
  volume components of multiple noise vectors setting the remaining components to zero. An example is 
  spin dilution where the first noise vector has non zero entries only on the first spin component, 
  the second vector only on the second spin component and so on. In this example, in order for the
  conditions in Eq.~(\ref{Eq:Noise_Condition}) to be satisfied, the total number of noise vectors 
  $N_r$ in the ensemble is restricted to multiplets of four. In Fig.~\ref{Fig:Dilution} we show
  a schematic representation of n-fold dilution. 
  \begin{figure}[!ht]
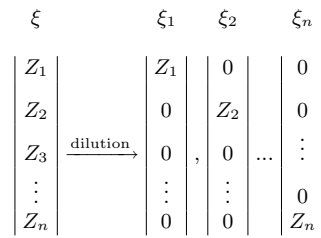

    \scalebox{0.9}{
      \mbox{$
        \begin{array}{|c|c|c|c|c|c|c|}
          \multicolumn{1}{c}{\xi}&\multicolumn{1}{c}{\,}&\multicolumn{1}{c}{\xi_1}&\multicolumn{1}{c}{\,}&\multicolumn{1}{c}{\xi_2}&\multicolumn{1}{c}{\,}&\multicolumn{1}{c}{\xi_n}  \\
          \multicolumn{7}{c}{\,}\\
          Z_1  &                       & Z_1  &          & 0   &         & 0\\
          Z_2  &                       & 0   &          & Z_2  &\phantom{\vdots} & 0\\
          Z_3  &\xrightarrow{\rm{dilution}}          & 0   &,          & 0   &...       & \vdots\\
          \vdots &                       & \vdots &          & \vdots &         & 0\\
          Z_n  &                       & 0   &          & 0   &         & Z_n
        \end{array}
        $}
    }
    \caption{A schematic representation of n-fold dilution. $Z_i$ denotes a random complex number.}
    \label{Fig:Dilution}
  \end{figure}
  
  The more one dilutes, the closer an estimate one gains of the all-to-all propagator. 
  This can be understood if one considers the extreme case where a noise vector is diluted 
  over all color, spin and volume components. In this case one would have inverted for each color, 
  spin and volume index thus obtaining the exact all-to-all propagator.
  
  The straight forward way to carry out the computation of the density-density correlator is 
  to expand Eq.~(\ref{Eq:DD_lattice}) on the quark level and replace each all-to-all 
  propagator with the stochastic average over the product between solution 
  and noise vectors: $G_{\mu\nu}^{ab}(x;y) = \langle\phi^a_\mu(x)\xi^{\dagger b}_\nu(y)\rangle_r$. 
  Throughout this paper we will refer to this as the direct method. 
  As demonstrated in Section IV,
  a reasonable estimate of the all-to-all propagators
  can be computed through the direct method if a large enough number of stochastic inversions is carried out. 
  
  Significant improvement to the results obtained using the direct method
  is achieved by applying the so called one-end trick.
  The one-end trick was
  originally devised to compute pion two-point functions~\cite{McNeile:2006bz}. 
  In its original form it is based on the realization that by appropriately combining solution vectors 
  one can derive the pion two-point function summed over both ends (source and sink).
  To be specific, let us consider the pion two-point function which, at the propagator level, is just the trace 
  of the absolute square of the quark propagator: 
  
  \be
  \sum_{\vec{x}}\langle\pi(\vec{x},t)|\pi(\vec{x}_0,t_0)\rangle = \sum_{\vec{x}}Tr\left[|G(\vec{x},t;\vec{x}_0,t_0)|^2\right].
  \label{Eq:Pi_2pt}
  \ee

  Let us consider the stochastic average over the product between two solution vectors given by:
  \be 
  \sum_{\vec{x}}\langle\phi^*(\vec{x},t;t_0)\phi(\vec{x},t;t_0)\rangle_r,
  \label{pion}
  \ee
  where the $t_0$ 
  appearing in the argument 
  of the solution vector is to indicate that the noise vectors are localized on this time slice, i.e:
  {\everymath{\displaystyle}
    \be
    \xi_\mu^a(\vec{x},t)_r = \xi_\mu^a(\vec{x})_r\delta(t-t_0),
    \ee 
    and hence
    \be
    \phi^a_\mu(\vec{x},t;t_0)_r = \sum_{\vec{y}}G^{ab}_{\mu\nu}(\vec{x},t;\vec{y},t_0)\xi_\nu^b(\vec{y})_r.
    \ee
  }
  By substituting for $\phi^a_\mu$ in Eq.~(\ref{pion}) we obtain:
  {\everymath{\displaystyle}
    \bea
    &\sum_{\vec{x}}\langle\phi^{*a}_\mu(\vec{x},t;t_0)\phi^a_\mu(\vec{x},t;t_0)\rangle_r =&\nonumber\\
    &\sum_{\vec{x},\vec{x}_0^\prime,\vec{x}_0^{\prime\prime}} G^{*ab}_{\mu\nu}(x;x_0^{\prime})G^{ac}_{\mu\kappa}(x;x^{\prime\prime}_0)
    \langle\xi^{*b}_\nu(x^{\prime}_0) \xi^{c}_\kappa(x^{\prime\prime}_0)\rangle_r =&\nonumber\\
    &\sum_{\vec{x},\vec{x}_0^\prime,\vec{x}_0^{\prime\prime}} G^{*ab}_{\mu\nu}(x;x^\prime_0)G^{ac}_{\mu\kappa}(x;x^{\prime\prime}_0)
    \delta_{bc}\delta_{\nu\kappa}\delta(\vec{x}^{\prime}_0-\vec{x}^{\prime\prime}_0)=&\nonumber\\
    &\sum_{\vec{x},\vec{x}_0^\prime}Tr\left[\left|G^{ab}_{\mu\nu}(x;x_0^\prime)\right|^2\right]&
    \eea}where $x_0^\prime = (t_0,\vec{x}_0^\prime)$ and $x_0^{\prime\prime} = (t_0,\vec{x}_0^{\prime\prime})$. 
  This is the pion two-point function given in Eq.~(\ref{Eq:Pi_2pt}) summed over all spatial source 
  and
  sink coordinates. This double summation increases statistics by spatial volume as compared to the standard way 
  where one computes two-point functions using a point-to-all propagator. The increase 
  by spatial volume 
  in statistics far outweighs the stochastic noise introduced by
  the stochastic inversion.
  
  The pion two-point function is the simplest implementation of the one-end trick since the $\gamma$-structure 
  of the interpolating fields combined with the backward propagator 
  of the antiquark yield a simple trace over a product of two forward quark propagators. To apply 
  the trick on an arbitrary meson two-point function with interpolating operators of the form
  $\bar{q}_f \Gamma q_{f^\prime}$, where $f \ne f^\prime$ label two flavors of quarks, not necessarily degenerate
  and $\Gamma$ an arbitrary combination of gamma matrices, one must use spin dilution. More explicitly, 
  the noise vectors should be of the form $\xi^a_\mu(x)_{(r,\sigma)} = \xi^a(x)_r \delta_{\mu \sigma}$. 
  The $r$ index counts sets of noise vectors, each set containing four noise vectors carrying an index 
  $\sigma$.
  We note here that this form of dilution is different than that described in the previous section. 
  Here the Z(2) random numbers involved in the spin dilution are the same for each spin component entry. 
  It can be easily confirmed that this choice satisfies the conditions 
  in Eqs.~(\ref{Eq:Noise_Condition}); the sum over the stochastic ensemble now becomes a double sum (over $r$ and $\sigma$) and $\langle\xi^a(x) \xi^{\dagger a^\prime}(x^\prime)\rangle_r=\delta(x-x^\prime) \delta_{a a^\prime}$.
  Within this notation the solution vectors are denoted as 
  $\phi^a_\mu(x)_{(r,\sigma)} = \sum_{x_0} G_{\mu\nu}^{ab}(x;x_0)\xi^b(x_0)_r\delta_{\sigma\nu}$. 
  Now one can appropriately combine the solution vectors to incorporate the $\Gamma$ matrices
  involved and obtain the meson two-point function summed over both ends:
  {\everymath{\displaystyle}
    \bea
    &\sum_{\vec{x},r} \phi^a_\mu(\vec{x},t;t_0)_{(r,\nu)} \Gamma^\prime_{\nu \sigma} \phi^{*a}_\kappa(\vec{x},t;t_0)_{(r,\sigma)}\bar{\Gamma}_{\kappa\mu}^\prime=&\nonumber\\
    &\sum_{\vec{x},\vec{x}_0^\prime,\vec{x}_0^{\prime\prime}} 
    G^{ab}_{\mu\nu}(x;x_0^\prime)\Gamma_{\nu \sigma}^\prime
    G^{*ab^\prime}_{\kappa\sigma}(x;x_0^{\prime\prime})\bar{\Gamma}_{\kappa\mu}^\prime \delta(\vec{x}_0^\prime - \vec{x}_0^{\prime\prime})\delta_{b b^\prime}=&\nonumber\\
    &\sum_{\vec{x},\vec{x}_0^\prime} Tr\left[
      G(x;x_0)\Gamma G(x_0;x)\bar{\Gamma}
      \right]&
    \eea
  }\noindent where $\Gamma^\prime = \Gamma\gamma_5$ and $\bar{\Gamma} = \gamma_0\Gamma^\dagger\gamma_0$. 
  Thus the one-end trick can be generalized to an arbitrary meson interpolating field. We would like 
  to note here that the automatic summation over the source using the same set of solution vectors
  selects a given momentum. Therefore the one-end trick by construction computes only two-point 
  functions at a specific momentum. In the examples given above this momentum was set to zero.
  To compute meson 
  two-point functions at various momenta, one must invert for a new set of solution 
  vectors having previously transformed the noise vectors with an appropriate phase. 
  In other words, one needs a new 
  set of stochastic inversions for each momentum vector.
  
  The crucial point that makes the one-end trick applicable to the evaluation of density-density
  correlators is the fact that the initial and final states have zero momentum.
  To show how to implement the one-end trick we consider the density-density correlator
  for an arbitrary meson with an interpolating operator of the form $\bar{q}_f \Gamma q_{f^\prime}$, where $f \ne f^\prime$:
  \bea
  C(\vec{x}_2)=\sum_{\vec{x}_1,\vec{x}} Tr\left[\gamma_5\gamma_0 G(x_1;x_0)\bar{\Gamma}^\prime G^\dagger(x_{2+1};x_0)\right.&\times&\nonumber\\
    \left.\gamma_5\gamma_0 G(x_{2+1};x)\Gamma^\prime G^\dagger(x_1;x) \right]&&
  \label{Eq:DDCorr_Prop}
  \eea
  where $x_{2+1} = (t_1,\vec{x}_2 + \vec{x}_1)$, $x_0 = (t_0,\vec{x}_0)$, $x_{1} = (t_1,\vec{x}_1)$, $x = (t,\vec{x})$ and $\Gamma^\prime = \Gamma\gamma_5$. 
  Let us define:
  \be
  S_{\mu\nu}^{ab}(\Gamma;x;y;t_0)\equiv \sum_r\phi^a_\mu(x;t_0)_{(r,\sigma)}\Gamma_{\sigma\kappa}\phi^{*b}_\nu(y;t_0)_{(r,\kappa)}
  \label{Eq:Def_S}
  \ee
  where $x = (t_x,\vec{x})$ and $y = (t_y,\vec{y})$ and the $t_0$ appearing in the argument of $S^{ab}_{\mu\nu}$
  is to indicate that the noise 
  vectors are localized on time-slice $t_0$. Summation over all repeated indices is implied.
  Assuming that the noise vectors are spin diluted in the manner described previously, we obtain
  
  \begin{align}
    S_{\mu\nu}^{ab}(\Gamma;x;y;t_0)&= \nonumber\\
    \sum_{\vec{x}_0,\vec{y}_0} G_{\mu\sigma}^{aa^\prime}(\vec{x},t_x;&\vec{x}_0,t_0)\Gamma_{\sigma\kappa}G_{\nu\kappa}^{*bb^\prime}(\vec{y},t_y;\vec{y}_0,t_0)
    \delta_{a^\prime b^\prime}\delta(\vec{x}_0-\vec{y}_0)\nonumber\\
    =\sum_{\vec{x}_0}G(&\vec{x},t_x;\vec{x}_0,t_0)\Gamma \left.G^\dagger(\vec{y},t_y;\vec{x}_0,t_0)\right|_{\mu\nu}^{ab}.
  \end{align}
  Thus in terms of the propagator defined in Eq.~(\ref{Eq:Def_S}), the expression
  \be
  \sum_{\vec{x}_1} Tr\left[\gamma_5\gamma_0S(\bar{\Gamma}^\prime;x_1;x_{2+1};t_0)\gamma_5\gamma_0S(\Gamma^\prime;x_{2+1};x_1;t)\right]
  \ee
  yields the density-density correlator of Eq.~(\ref{Eq:DDCorr_Prop}) with an additional summation 
  over the source coordinate $\vec{x}_0$. 
  This is the generalization of the one-end trick to meson four-point correlators. It is apparent that 
  one needs two sets of stochastic inversions: one with the noise vectors localized on the source 
  time-slice $t_0$ and one with the noise vectors localized on the sink time-slice $t$. 
  
  \section{Interpolating fields and Lattice parameters}
  For the pion and the $\rho$-meson we compute the density-density correlators using both
  the one-end trick and the direct method.
  For the nucleon and the $\Delta$ it is 
  not as straight forward to apply the one-end trick. The quark
  line propagating without a density insertion
  complicates the generalization of the trick to baryons since the 
  propagators to be replaced
  by noise vectors are odd in number and therefore unlike for mesons
  the noise vectors cannot be grouped in pairs to yield $\delta$-functions
  after summation. 
  Thus in this work for the nucleon and $\Delta$ density-density correlators
  we only present results using the
  direct method.
  
  One of our main goals is to detect a possible asymmetry in the charge distributions of these particles.
  For this purpose we select interpolating operators so that they project to physical spin states.
  For the mesons we use interpolating operators of the form: $J^M = \bar{u}\Gamma d$
  with $\Gamma = \gamma_5$
  for the case of the pion and 
  $\Gamma = \{\frac{\gamma_1 - i\gamma_2}{2},\gamma_3,\frac{\gamma_1 + i\gamma_2}{2}\}$
  for the $+1$, $0$ and $-1$ polarizations of the vector meson respectively, where
  we have taken the $z$ axis
  along the spin axis. For the nucleon we use $J^N_\sigma = \epsilon^{abc}u_\sigma^a(u^{b\top}C\gamma_5d^c)$
  where $C=\gamma_0\gamma_2$. For the case of the $\Delta$ we opt to probe the spin $\pm \frac{3}{2}$ 
  components. Thus we use the interpolating operators: 
  \be
  \begin{array}{rcl}
    J^\Delta_{+\frac{3}{2}} = &\frac{1}{\sqrt{3}}\epsilon^{abc}\left[u^a_1(2u^{b\top}C\Gamma_+d^c)+d_1^a(u^{b\top}C\Gamma_+u^c)\right]\\
    &\\
    J^\Delta_{-\frac{3}{2}} = &\frac{1}{\sqrt{3}}\epsilon^{abc}\left[u^a_2(2u^{b\top}C\Gamma_-d^c)+d_2^a(u^{b\top}C\Gamma_-u^c)\right]
  \end{array}
  \ee
  where $\Gamma_\pm = \left(\gamma_1 \mp i\gamma_2\right)/2$.

  Given the large number of inversions needed to compute the density-density correlators and the available computer resources,
  using dynamical Wilson fermions that are
  fast to invert is the only option at our disposal.
  We use two dynamical degenerate flavors of Wilson fermions
  at three pion masses.
  The exact parameters of the ensembles used are listed in Table~\ref{Table:Confs}.
  
  \begin{table}[h]
    \caption{The first column gives the number of configurations analyzed, the second the value of
      the hopping parameter, the third the pion mass in GeV, the fourth the ratio of the pion mass 
      to the $\rho$ mass, the fifth 
      the nucleon mass in GeV and the last column the size of the lattice.
      The first two sets of configurations are from Ref.~\cite{Orth:2005kq} while the third is from Ref.~\cite{Jansen:2005yp}.
      The lattice spacing is determined from the nucleon mass at the chiral limit.}
    \label{Table:Confs}
    \begin{tabular}{cccccc}
      \hline\hline
      \multicolumn{6}{c}{$\beta = 5.6$, $a^{-1}=2.56(10)$~GeV}\\
      $N_{\rm conf}$ & $\kappa$ & m$_\pi$ (GeV) & m$_\pi$/m$_\rho$ & M$_N$ (GeV)& L$^3\times$T \\
      \hline
      185 & 0.1575\phantom{0} & 0.691(8) & 0.701(9)\phantom{0} & 1.485(18) & 24$^3\times$40\\
      150 & 0.1580\phantom{0} & 0.509(8) & 0.566(12)   & 1.280(26) & 24$^3\times$40\\
      200 & 0.15825           & 0.384(8) & 0.453(27)   & 1.083(18) & 24$^3\times$32\\
      & $\kappa_c$ = 0.1585 & 0  &      & 0.938(33) & \\
      \hline\hline
    \end{tabular}
  \end{table}
  
  To suppress excited state contributions we use Gaussian or Wuppertal smeared sources \cite{Alexandrou:1992ti}.
  In addition we apply
  hypercubic (HYP) smearing~\cite{Hasenfratz:2001hp}
  on the gauge links that enter the smearing function that builds the Gaussian smearing function. 
  The parameters that enter the Gaussian smearing function are taken
  from Ref.~\cite{Alexandrou:2006ru} where they were determined by optimizing ground state dominance for 
  the nucleon. In fact, in Ref. \cite{Alexandrou:2006ru} it was demonstrated that one can
  damp excited state contributions to the nucleon two-point function as early as 0.3~fm from the 
  source time slice.
  The parameters for the HYP smearing are taken 
  from Ref.~\cite{Hasenfratz:2001hp}.
  
  For the computation of the correlators we take the time-slice of the density insertions
  to be at mid-point of the time separation between sink and source. For the direct method we
  take the time separation 
  between the sink and the source to be $t -t_0= 10a$ or 0.77~fm. This is the minimum time
  separation that is needed for the suppression of excited states.
  For the one-end trick the separation 
  between sink and source is set to $t- t_0 = 14a$.
  The reason for taking a larger
  time separation when using the one-end trick lies in the
  accuracy of the results that allows for a larger time separation with a good signal.
  This allows us to check that indeed excited state contributions are sufficiently suppressed
  by comparing results at the two sink-source time separations. 
  
  We first give the details of the computation in the case of the direct method. 
  We require two sets of stochastic 
  propagators per configuration,
  one with the noise vectors localized on the insertion time-slice and one with the noise vectors
  localized on the sink. We also compute a point-to-all propagator from the source 
  time-slice to all lattice sites.
  The noise vectors are diluted in color, spin and even-odd spatial sites. 
  Dilution in time is automatic here since we invert with the noise vectors localized on a single time-slice.
  Thus each noise vector is diluted to twenty-four independent noise vectors 
  requiring twenty-four times more inversions. 
  The number of noise vectors used is determined through a tuning process. 
  For this tuning the $\Delta$-baryon correlator at the lightest pion mass is considered.
  By comparing the decrease of the relative statistical error when increasing on one hand statistics and
  on the other hand the number of noise vectors used,
  we determine the optimum number of stochastic vectors.
  For this tuning we use 50 configurations and 
  compute the $\Delta$-baryon correlator for three, six and nine such 24-fold 
  diluted noise vectors. For $N_r$=3, 6 and 9 we find a relative statistical error of 50\%, 
  20\% and 16\% respectively. The fact that by doubling the number of noise vectors from 3 to 6
  the statistical error decreases by more than one half is an indication that $N_r$=3 
  is too small yielding large stochastic noise. 
  On the other hand, increasing the number of noise 
  vectors from 6 to 9 the relative error decreases by $\sqrt{6/9}$, which is what is expected
  from scaling. This indicates that at this point increasing $N_r$ or the 
  number of configurations is equivalent.
  We thus fix the number of noise vectors to six.
  Since we carry out two sets of stochastic inversions, one at the sink and one at the
  insertion time-slice, and since we use color, spin and even-odd dilution 
  we need 288 stochastic inversions per 
  configuration. 
  This amounts to a total of 300 inversions per density-density correlator 
  if we additionally consider the 
  point-to-all propagating from the origin. 
  To increase statistics for the two ensembles 
  corresponding to the two lightest pion masses needed for the baryons, 
  we calculate density-density correlators using the first and second 
  half time interval of each configuration. 
  Furthermore, for the lightest pion mass we improve 
  statistics by using $N_r$=9 noise vectors for the correlators.
  Thus for $\kappa=0.1580$ we carry out 600 
  inversions per configuration while for $\kappa=0.15825$ 888 inversions per configuration.
  
  For the case of the mesons we have additionally computed the 
  charge distributions using the one-end trick. 
  Therefore for the computation of the meson density-density correlators additional inversions 
  are carried out since the dilution method is specific to the one-end trick. Like for the direct method,
  two sets of inversions are needed to extract the density-density correlator using the one-end trick: 
  one set with the noise source 
  set at the source time $t_0$ and one set with the noise source set at the sink time $t$. 
  We use eight spin-diluted noise vectors amounting to 32 inversions at the source and 32 at the sink
  or a total of 64 per configuration.
  
  \section{Results}
  \subsection{Comparison between the direct method and the one-end trick}
  For the meson density-density correlators we can compare results obtained using the direct method
  with those using the one-end trick. Given that the time separation between sink and source is
  larger in the latter case this also provides a check of ground state dominance.

  \begin{figure}[!h]
    \includegraphics[width=\linewidth]{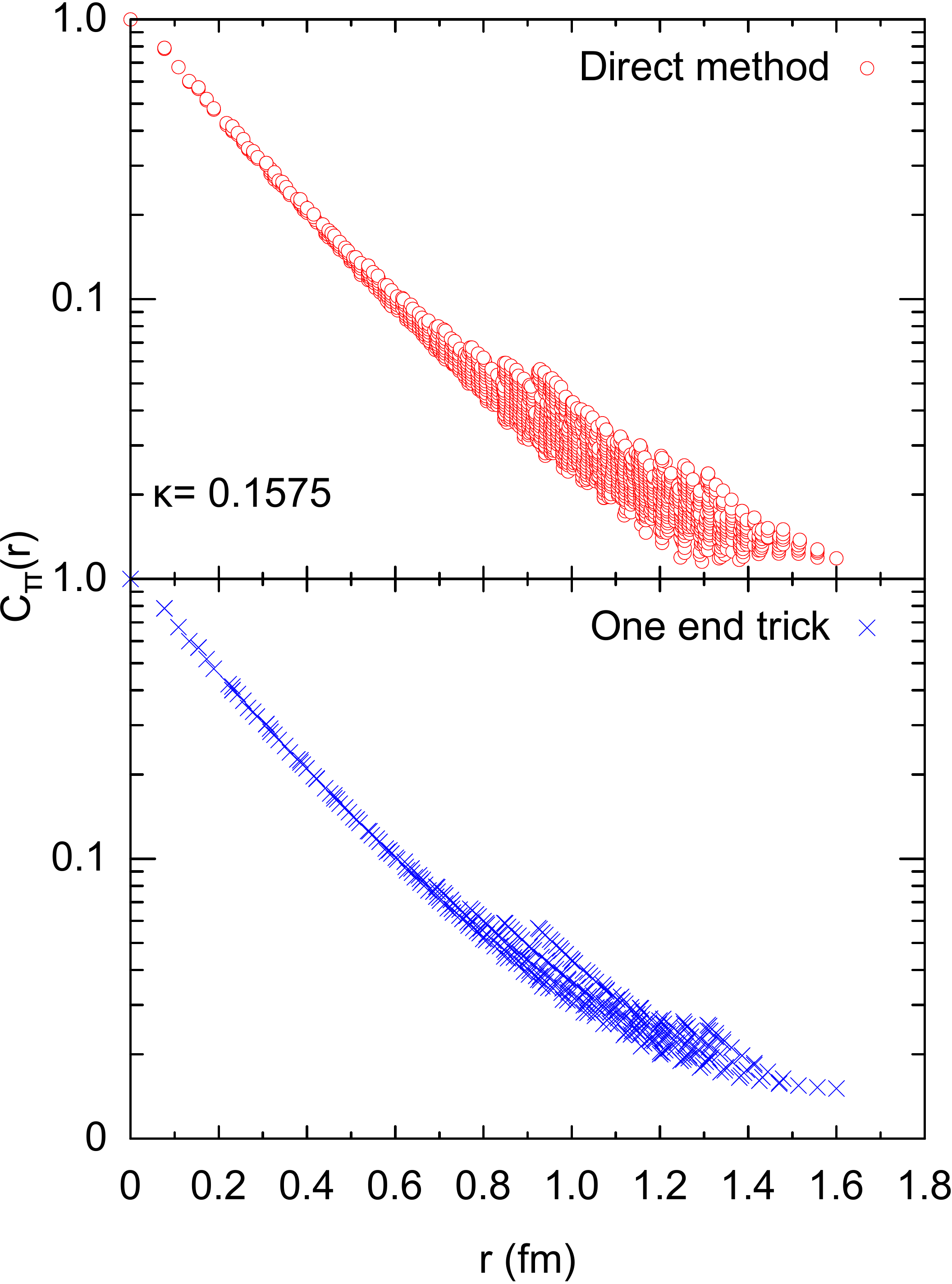}
    \caption{The pion density-density correlator using the one-end 
      trick (upper graph) and using the direct method (lower graph).
      The mean value of $C_\pi(r)$ is plotted as described in the text and error bars are suppressed for clarity.}
    \label{Fig:OE_DM_cmp_pion}
  \end{figure}

  \begin{figure}
    \includegraphics[width=\linewidth]{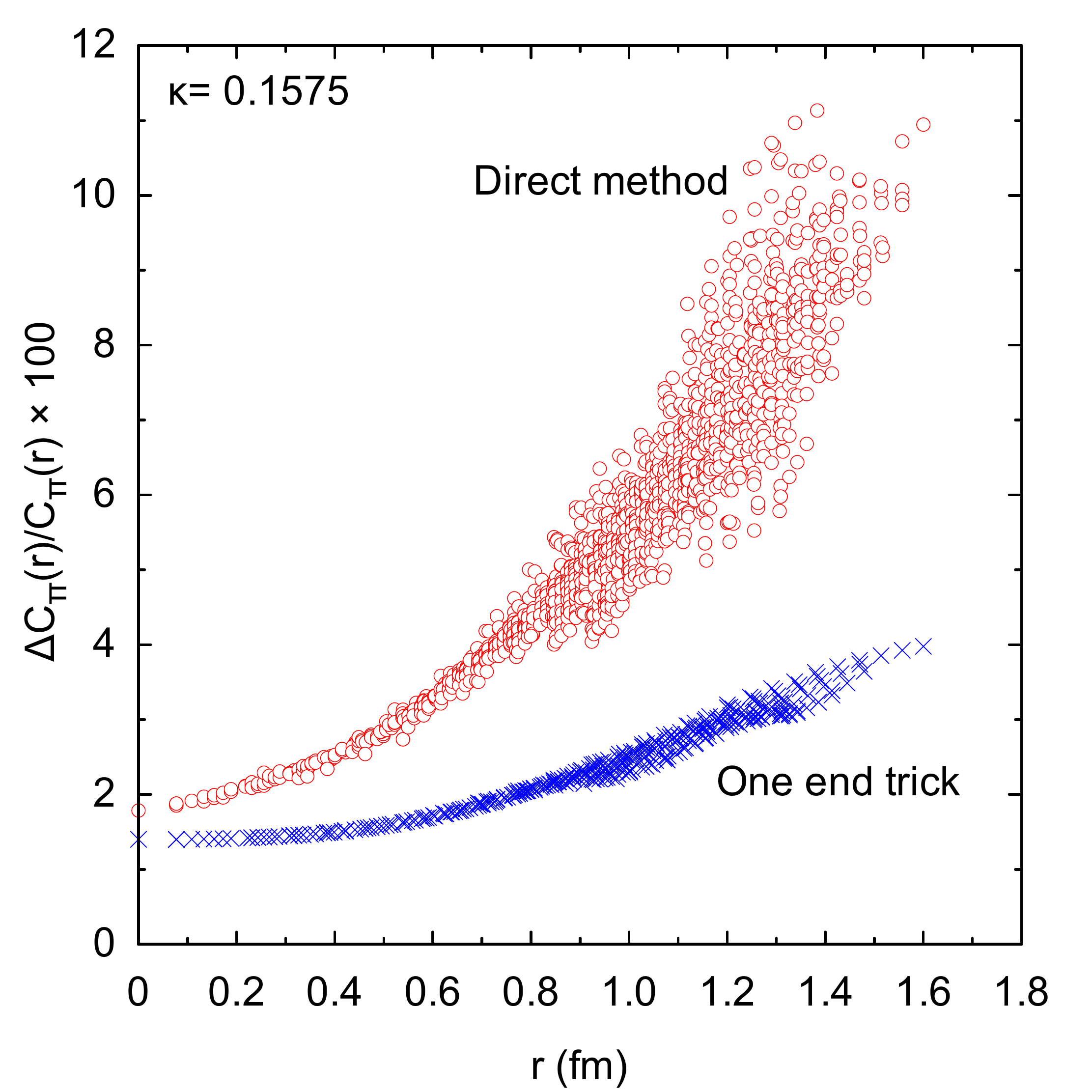}
    \caption{Comparison between the relative error of the correlator computed with the one-end trick (blue crosses) and the direct method (red circles).}
    \label{Fig:OE_DM_errcmp_pion}
  \end{figure}
  
  The main source of error is due to the stochastic noise when computing the all-to-all propagators. By implementing the one-end trick, the four-point function is 
  automatically summed over sink and source coordinates and thus this method is expected to suppress 
  stochastic noise considerably. 
  \begin{figure}
    \includegraphics[width=0.95\linewidth]{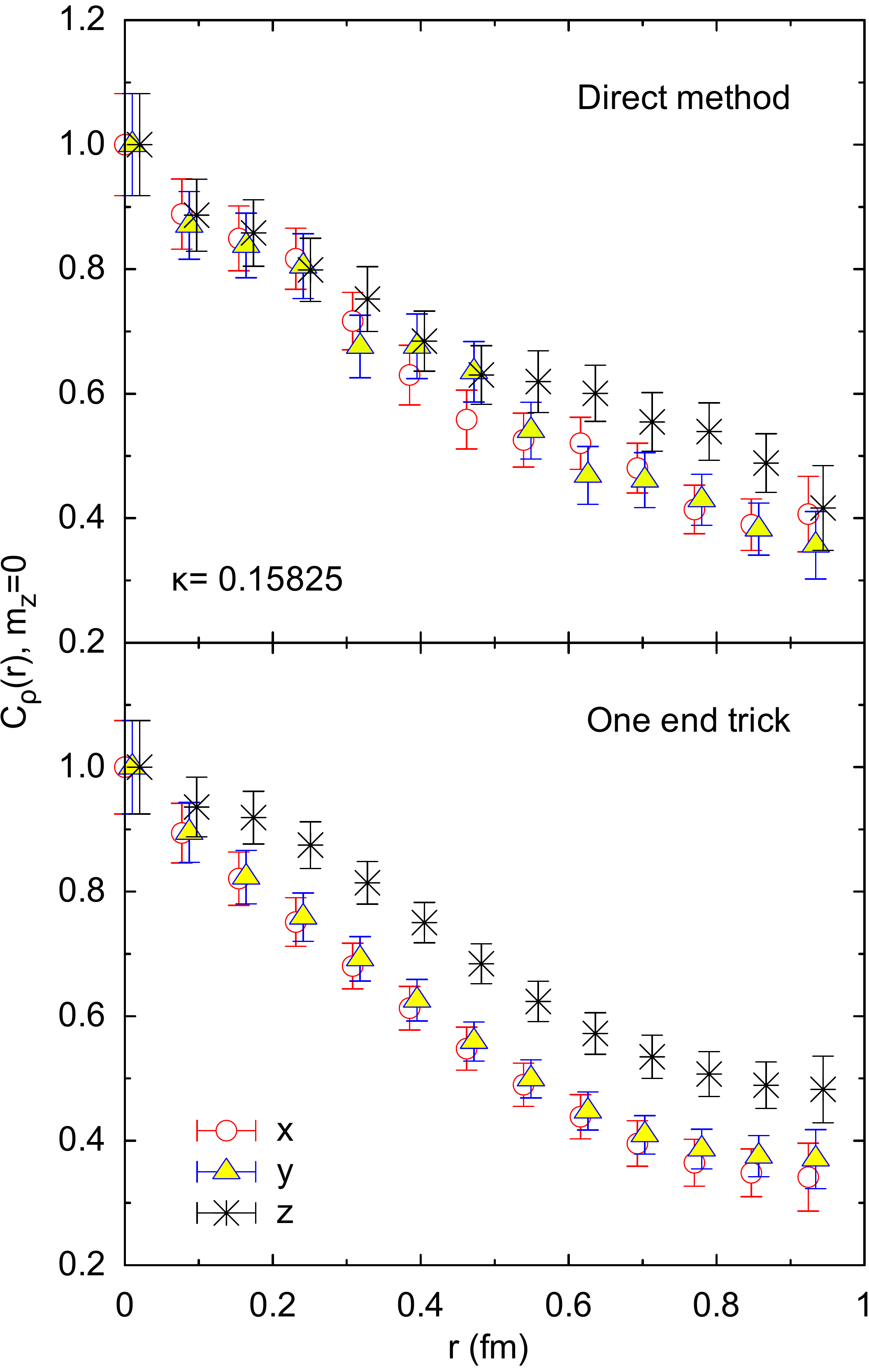}
    \caption{Comparison between the vector meson $m_z=0$ correlator projected along the three axes computed with the direct method (upper graph) and with the one-end trick (lower graph) using 200 configurations.}
    \label{Fig:OE_DM_cmp_rho0}
  \end{figure}
  
  In Fig.~\ref{Fig:OE_DM_cmp_pion} we show the pion correlator computed using the one-end trick 
  and the direct method as a function of the distance from the origin. To avoid having to display 
  all lattice points in the graph we replace points lying within a cell of size 0.015~fm$\times$0.05
  by their average. 
  We normalize the correlator by dividing by its value at the origin.
  The errors in Fig.~\ref{Fig:OE_DM_cmp_pion} are not shown for clarity. 
  As can be seen, we find that the two methods yield consistent 
  results for the correlators. This demonstrates that excited states are
  sufficiently suppressed with a sink-source separation of 10 time slices.
  However, at a given distance $r$, 
  the correlator computed using the direct method shows more spread than the one computed using the one-end 
  trick. That this reflects larger statistical noise is shown in
  Fig.~\ref{Fig:OE_DM_errcmp_pion}, where we compare the relative errors of the two binned 
  correlators. As can be seen, at large distances the maximum relative error exhibited by the one-end trick method 
  is around 4\% while for 
  the direct method exceeds 10\%. This is a direct consequence 
  of the double sum accomplished with the implementation of the one-end trick.
  In addition, when using the one-end trick the density-density correlator of a state of spin projection $m_z=0$ 
  is symmetric under reflections of the spatial coordinates i.e. $C(\vec{r})=C(-\vec{r})$ by construction 
  whereas in the direct method it is symmetric only statistically. 
  For the $m_z=\pm1$ projections of the vector meson we instead have $C^{m_z=+1}(\vec{r})=C^{m_z=-1}(-\vec{r})$.
  Because of this symmetry we average over the results for the $m_z=+1$ and $m_z=-1$ spin projections 
  and hereby denote this correlator by $m_z=\pm1$.
  The same is done for the spin projections $m_z=\pm 3/2$ of the $\Delta$. 
  The reduction of the error by more than a factor two when 
  using the one-end trick
  comes at a reduced computational cost. In the one-end trick the 
  computation of the 
  correlator is done using 64 inversions while
  for the direct method used in this comparison 
  we carried out 300 inversions per configuration i.e. we need
  4.7 times less inversions for twice the accuracy.
  This, combined with the fact that the computation using the one-end trick 
  is carried out for a source-sink separation of 14 time slices while for the direct method we
  used a separation of 10 time slices and given that
  relative errors grow exponentially with the sink-source separation, clearly shows the superiority of
  the one-end trick.

  \begin{figure}[!h]
    \includegraphics[width=1\linewidth]{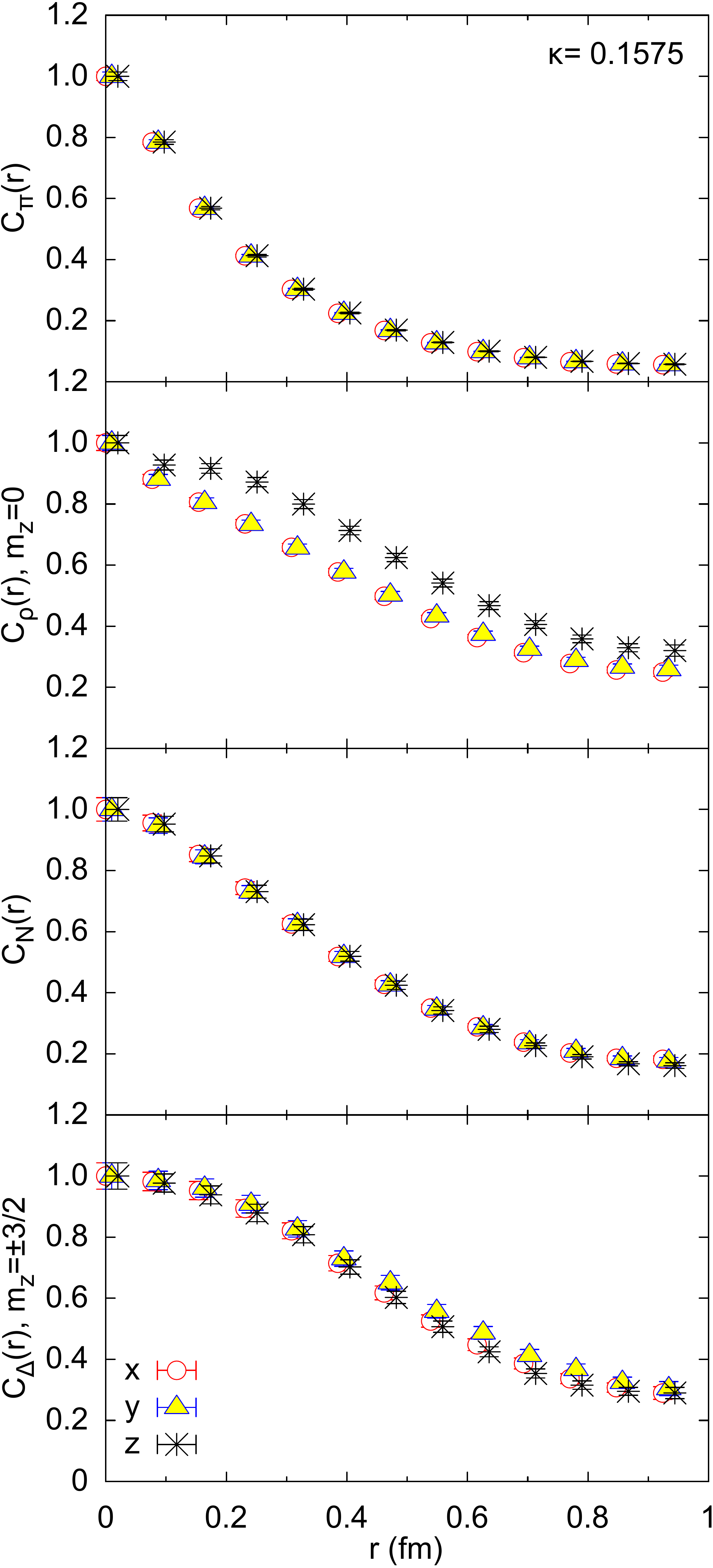}
    \caption{Projections of the correlator along the three axes.
      From top to bottom: For the pion and the $\rho$ - meson using the
      one-end trick with 64 inversions, 
      for the nucleon and the $\Delta$ using the direct method with
      300 inversion needed per configuration.}
    \label{Fig:axesp_k0p1575}
  \end{figure}

  One of the main goals of this calculation is to detect a possible asymmetry in the hadron charge distribution. 
  In Fig.~\ref{Fig:OE_DM_cmp_rho0} we compare the two methods for the case of the $m_z=0$
  spin projection of the vector meson at the lowest pion mass available using the same number of configurations.
  Only the profile of the correlator along the three axes is plotted so that we can detect a possible asymmetry. 
  As can be seen, an elongation 
  along the $z$ axis is clearly observed only when using the one-end trick.
  The statistical error in the 
  direct method is not small enough to draw definite conclusions, 
  since the projections of the correlator on the three axes 
  are within error bars. Using the one-end trick the fluctuations 
  are small enough to conclude that the vector meson 
  is indeed prolonged along the $z$ axis. 
  When discussing results on baryon deformation one 
  has to keep in mind that statistical fluctuations are larger
  than for mesons and that
  we can only apply the direct method making reaching conclusions for baryons
  more difficult. 
  
  Having demonstrated the effectiveness of the one-end trick 
  in suppressing stochastic noise, 
  all meson observables that we present hereon are 
  computed with the one-end trick.

  \subsection{Results without volume corrections}
  In Fig.~\ref{Fig:axesp_k0p1575} we show the density-density correlators
  for the pion and the spin zero projection ($m_z=0$) of the $\rho$-meson using the one-end trick and for
  the nucleon and spin $m_z=\pm \frac{3}{2}$ projection of the $\Delta$ using the direct method. 
  All correlators are projected along the three axes to display a possible asymmetry. 
  This is done for the smallest pion mass available, namely $m_\pi=0.691(8)$~GeV.
  As can be seen, a clear elongation of the vector meson along the $z$ axis is observed confirming
  our previous results~\cite{Alexandrou:2002nn}. The asymmetry is clearly smaller than for the lightest pion
  mass shown in Fig.~\ref{Fig:OE_DM_cmp_rho0}, showing that the deformation
  increases as the pion mass decreases. 
  On the other hand, the nucleon shows no asymmetry within this method. 
  For the $\Delta$ although there is a tendency for results
  projected along the $z$-axis to lie lower, all projections are well
  within error bars and therefore no asymmetry can be
  claimed. As pointed out when discussing results on the $\rho$
  using the direct method, statistical
  errors can hide possible deformation and one may have to improve
  on the errors to detect a small asymmetry.
  
  Another way to visualize the asymmetry is to construct
  two-dimensional contour plots. Fig.~\ref{Fig:cont_rho0_all} shows a contour plot of the $m_z=0$ spin $\rho$-meson state
  on the $x$ - $z$ plane. As can be seen, 
  the contours are elongated along the $z$-axis as compare to a circle of radius equal to
  the distance along the $x$-axis for all three pion masses showing a clear asymmetry.
  This leads to the 
  conclusion that the vector meson in the spin projection zero state is prolate. 
  On the other hand, the $m_z=\pm1$ $\rho$-meson state, shown in Fig.~\ref{Fig:cont_rho1_all} 
  shows the opposite behavior.
  Namely the correlator is found to be larger along the $x$-axis, as compared to a circle, evidence that in this
  spin state
  the $\rho$ is in fact an oblate. This is in agreement with what is derived in Ref.~\cite{Alexandrou:2002nn} where
  it is shown that if the spin-0 state is a prolate the $\pm 1$ channels will be oblate with about half the amount of deformation.
  The fact that the $\rho$-meson in its maximal spin projection state is an oblate is in agreement with a recent
  calculation of a negative electric quadrupole form factor evaluated in quenched lattice QCD~\cite{Hedditch:2007ex}. 
  
  \subsection{Results after finite volume corrections}
  Density-density correlators computed in a finite box with periodic
  boundary conditions are susceptible to finite volume effects. 
  Finite volume effects mostly affect the tail of the distributions and need to be corrected.
  To perform these corrections we follow the analysis developed in Ref.~\cite{Burkardt:1994pw}.
  The density-density correlation function computed 
  on a lattice of spatial 
  dimension $L$ can be written as an infinite sum over the Brillouin zones
  \begin{equation}
    C(\vec{r}) = \sum_{\vec{n}=0}^{\infty}C_0(\vec{r}+\vec{n}L)
    \label{Eq:images}\end{equation}
  where $C(\vec{r})$ is the density-density correlator
  computed on the periodic lattice and $C_0(\vec{r})$ is the ``correct'' correlator that
  one would compute if the lattice were of infinite size.
  Thus the correlation function computed in a finite box with periodic boundary conditions is in fact a sum 
  of all images arising from the surrounding boxes. 
  Since $C_0(\vec{r})$ is a fast decaying function, approximated by exponential or Gaussian dependence on the radius,
  it means that the 
  leading contributions to the sum come from the nearest neighboring Brillouin zones. 
  A two-dimensional sketch drawn in Fig.~\ref{Fig:images} demonstrates the images that contribute to the
  correlator. In this figure, 
  the asterisk shows the origin of the fundamental cell (white box) while the triangles show the origins of
  the neighboring cells (gray boxes). To first order, the correlator computed in the white box is 
  a superposition of the correlator with origin the asterisk and the eight correlators with origins the filled triangles, 
  in accord with the expression given in Eq.~(\ref{Eq:images}). 
  Thus the correlator that we compute on a periodic lattice is overestimated.
  This is particularly severe close to the boundaries of the lattice 
  where contributions from the images are largest. For example, the correlator at the distances indicated
  by the filled circles in Fig.~\ref{Fig:images} is approximately twice as large
  as the ``correct" correlator since besides the contribution from the fundamental cell,
  a neighboring cell contributes equally as indicated by the dashed line. Similarly, the correlator 
  computed at the distances indicated by the open circles at the corners of the fundamental cell 
  is approximately four times larger since there are contributions from
  three neighboring cells, as shown by the dotted line.
  
  This analysis can be extended to three dimensions. The correlator is twice as large at the six distances
  given by
  $\pm L/2\hat{n}_i,i={x,y,z}$ where $\hat{n}_i$ is the unit vector in the $i$-direction.
  Similarly, the correlator is four times as large at the twelve distances
  $L/2 (\hat{n}_i \pm \hat{n}_j),i\ne j$
  and eight times as large at the eight corners $L/2 (\pm \hat{n}_x \pm \hat{n}_y \pm \hat{n}_z)$.
  
  \bw
  
  \begin{figure}[!h]
    \includegraphics[width=0.9\linewidth]{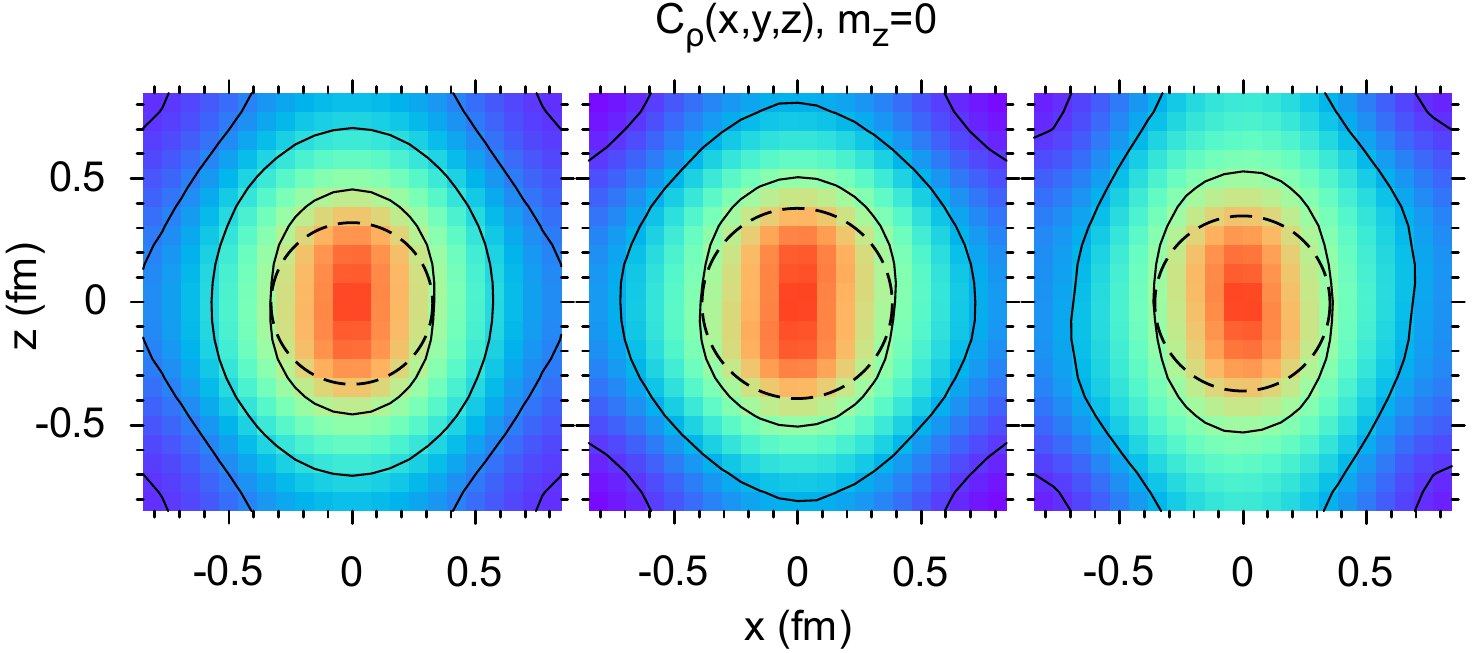}
    \caption{The correlator of the $\rho$ - meson, $m_z=0$ projected on the $x$ - $z$ plane for decreasing pion mass left to right. The dashed circles are to guide the eye.}
    \label{Fig:cont_rho0_all}
  \end{figure}
  \begin{figure}[!h]
    \includegraphics[width=0.9\linewidth]{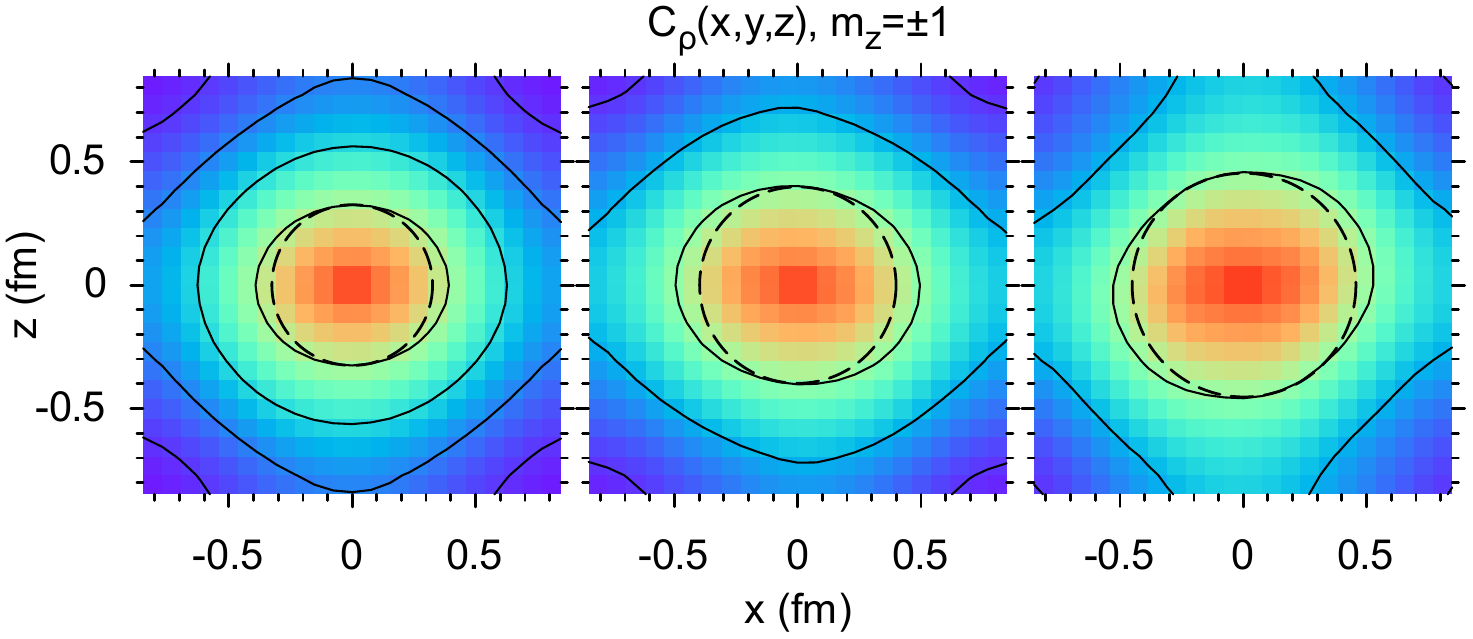}
    \caption{The correlator of the $\rho$ - meson, $m_z=\pm1$ projected on the $x$ - $z$ plane for decreasing pion mass left to right.
      The dashed circles are to guide the eye.}
    \label{Fig:cont_rho1_all}
  \end{figure}
  
  \ew
  
  \begin{figure}[h]
    \includegraphics[width = 0.87\linewidth]{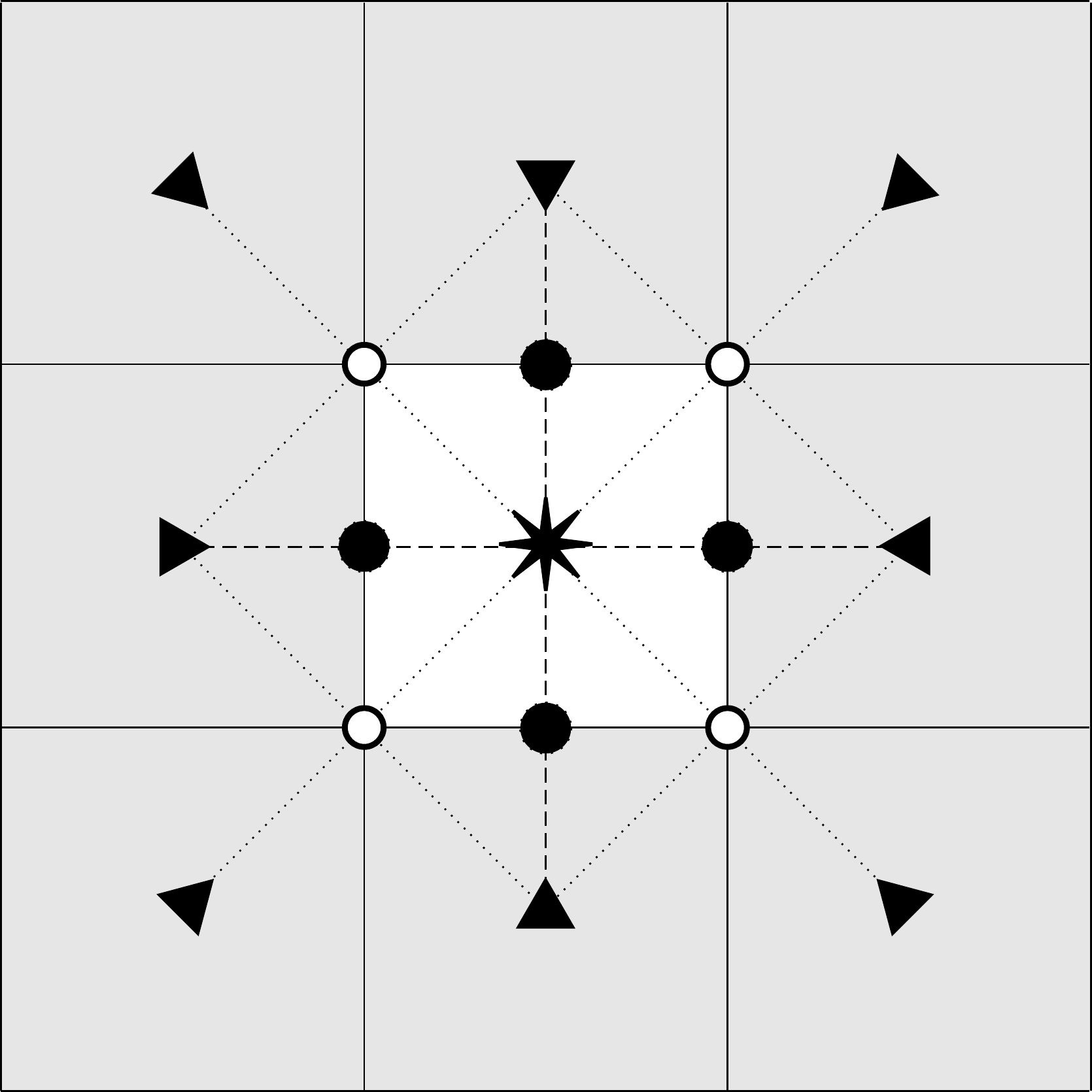}
    \caption{Two-dimensional example of image contributions. The correlator computed at the filled circles
      (open circles) is approximately two (four) times larger than the 
      ``correct" correlator.}
    \label{Fig:images}
  \end{figure}
  
  \begin{figure}[t]
    \includegraphics[width = 0.925\linewidth]{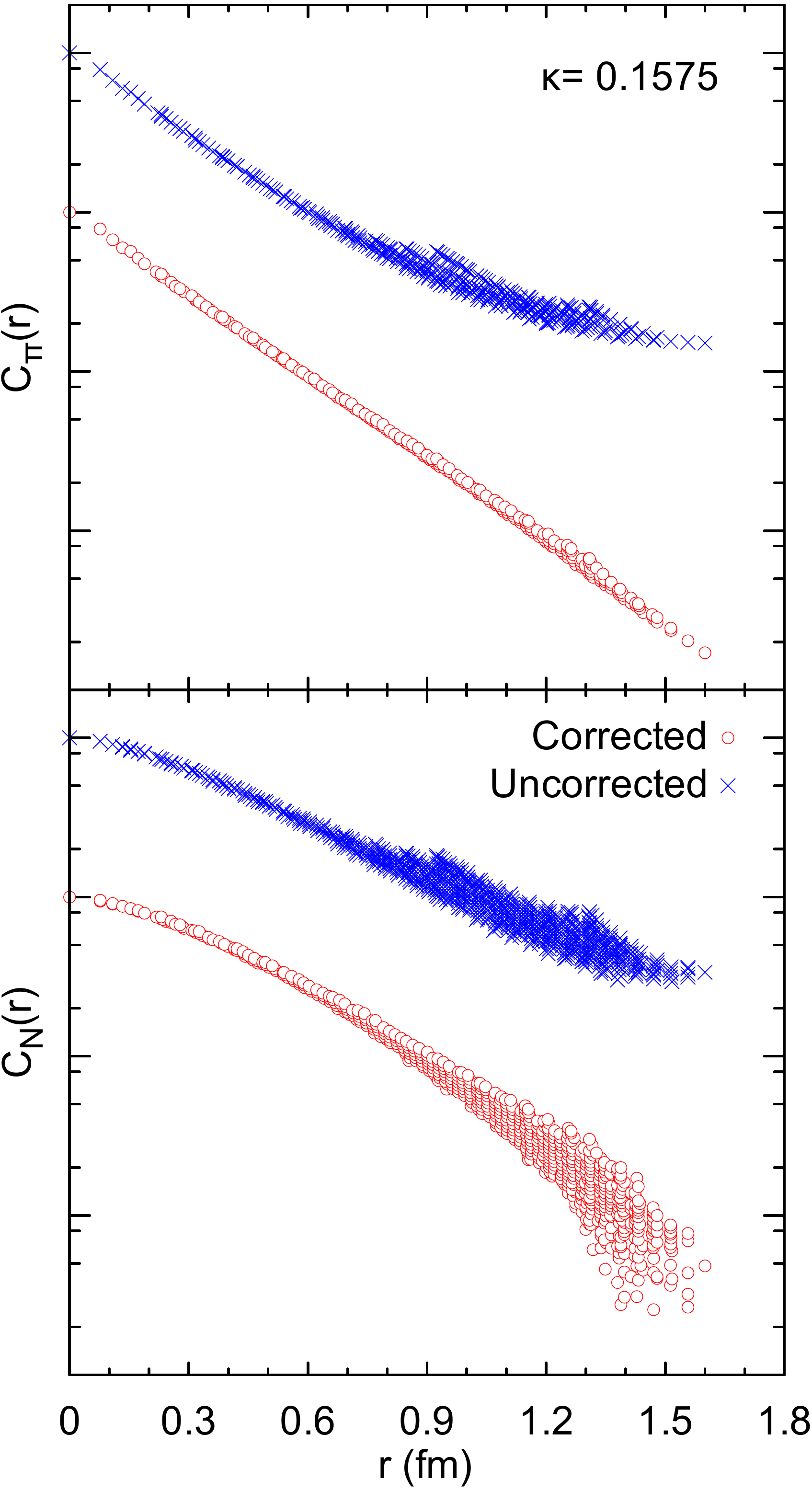}
    \caption{The pion correlator (top) and the nucleon correlator (bottom)
      as computed on the lattice (crosses) and corrected for the images of nearest neighboring 
      lattices (open circles).
      The corrected correlator is divided by a factor of ten for clarity.
      Data are binned and error bars are omitted to avoid cluttering.}
    \label{Fig:img_pn}
  \end{figure}

  \begin{figure}[t]
    \includegraphics[width = 0.925\linewidth]{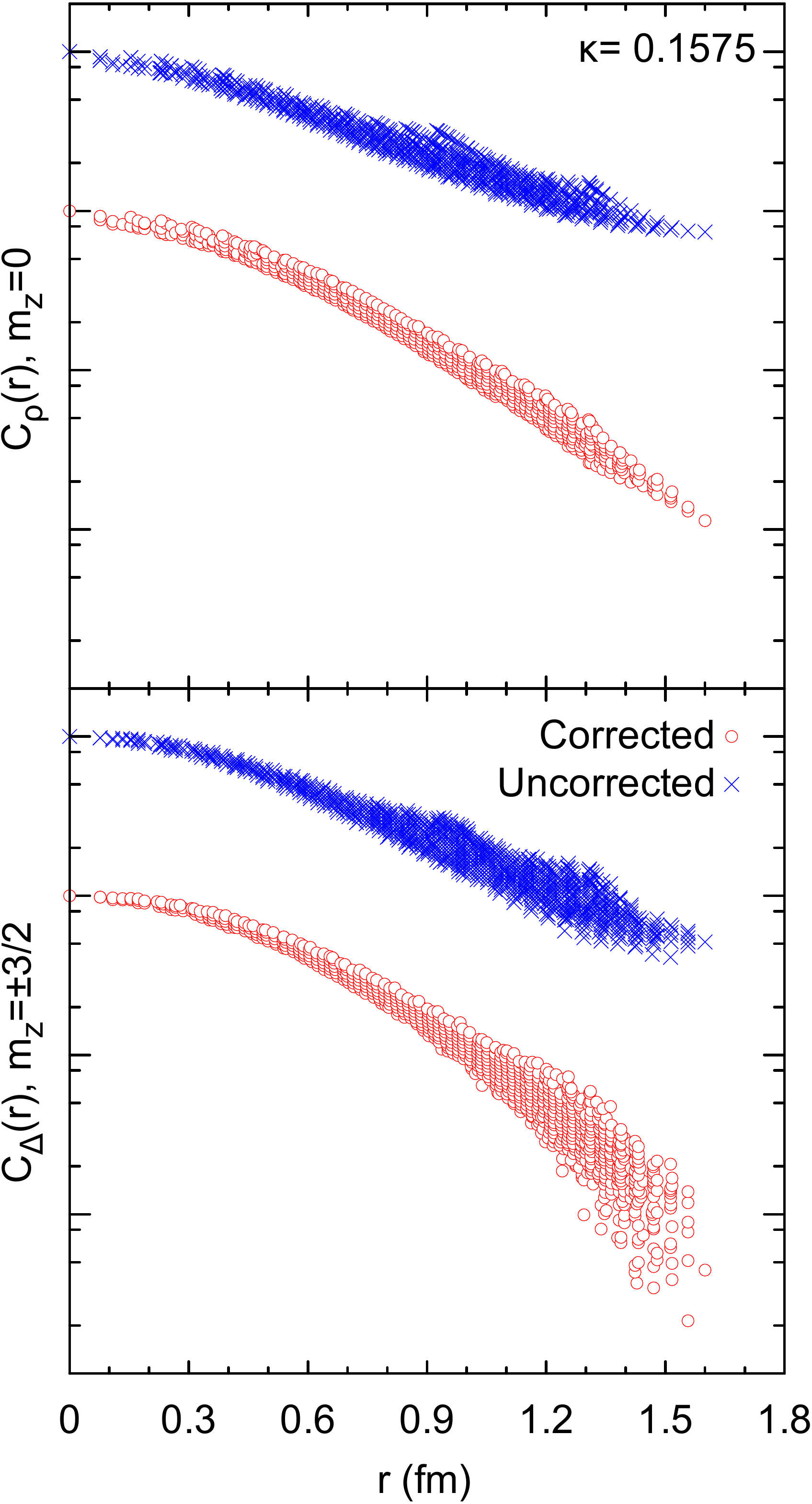}
    \caption{The $\rho$-meson, $m_z=0$ correlator (top) and the $\Delta$, $m_z=\pm3/2$ correlator (bottom). The notation is the same as that of Fig.~\ref{Fig:img_pn}.}
    \label{Fig:img_rd}
  \end{figure}

  \begin{table}[h]
    \caption{The parameters obtained from fitting the sum of images to the lattice data.}
    \label{Table:FitParams}
    \begin{tabular}{cr@{.}lr@{.}lr@{.}l} 
      \hline
      \hline
      \multicolumn{1}{c}{$\kappa$}&\multicolumn{2}{c}{0.1575} &\multicolumn{2}{c}{0.1580}&\multicolumn{2}{c}{0.15825}\\
      \hline
      \multicolumn{7}{c}{Mesons}\\
      \hline
      \multicolumn{7}{c}{$\pi$}\\
      $A_0$ &0&986(21) &1&129(33) & 1&437(78)\\
      $m_0$ &0&307(7) &0&405(11) & 0&579(25)\\
      $\sigma$&0&993(7) &0&884(9) & 0&779(12)\\
      \multicolumn{7}{c}{$\rho$, $m_z=0$}\\
      $A_0$ & 0&969(13) & 0&964(21) & 0&919(31) \\
      $m_0$ & 0&0173(19) & 0&0140(26) & 0&0093(26) \\
      $A_1$ & 0&00170(31) & 0&0031(16) & 0&00183(46) \\
      $m_1$ & 0&0466(87) & 0&077(33) & 0&0033(12) \\
      $\sigma$ & 1&615(41) & 1&646(69) & 1&76(11)  \\
      \multicolumn{7}{c}{$\rho$, $m_z=\pm1$}\\
      $A_0$ & 0&976(10) & 0&961(16) & 0&977(28) \\
      $m_0$ & 0&0194(16) & 0&0128(16) & 0&0141(34) \\
      $A_1$ & -0&00113(18)&-0&00054(34) &-0&0012(17) \\
      $m_1$ & 0&0560(91) & 0&025(12) & 0&066(69) \\
      $\sigma$ & 1&577(30) & 1&659(47) & 1&613(87) \\
      \hline
      \multicolumn{7}{c}{Baryons}\\
      \hline
      \multicolumn{7}{c}{$N$}\\
      $A_0$ & 1&014(39) & 1&039(34) &1&057(34) \\
      $m_0$ & 0&0673(40) & 0&0698(44) &0&0548(38)\\
      $\sigma$ & 1&451(20) & 1&413(22) &1&450(24) \\
      \multicolumn{7}{c}{$\Delta$, $m_z=\pm\frac{3}{2}$}\\
      $A_0$ & 1&024(22) & 1&033(19) & 1&023(16) \\
      $m_0$ & 0&0125(11) & 0&0130(12) & 0&0087(8) \\
      $A_1$ &-0&00029(25) &-0&0007(14) &-0&00121(49)\\
      $m_1$ & 0&024(13) & 0&022(25) & 0&077(30) \\
      $\sigma$ & 1&750(32) & 1&708(34) & 1&787(33) \\
      \hline
      \hline
    \end{tabular}
  \end{table}

  All results that have been discussed so far are for the correlators computed on the lattice 
  with no corrections applied 
  for the images. For the analysis of quantities, such as the root mean squared radius,
  that are sensitive to the long distance behavior of the
  distributions it is important to take in to account the image contributions and define a corrected 
  correlator. 
  To correct for the images and extract $C_0(\vec{r})$ of Eq.~(\ref{Eq:images}) by knowing only $C(\vec{r})$
  we need to have an Ansatz for the asymptotic behavior of $C_0(\vec{r})$. 
  If the asymptotic behavior is known then we can subtract from the lattice data
  the contribution from the images, up to a given order, and extract $C_0(\vec{r})$. 
  In this work, we consider only nearest neighbor contributions to the correlator. Thus Eq.~(\ref{Eq:images}) 
  becomes:
  
  \begin{equation}
    C(\vec{r})\simeq\sum_{\left|\vec{n}\right|\in[0,\sqrt{3}]} C_0(\vec{r}+\vec{n}L).
    \label{Eq:img_1st}
  \end{equation}
  
  We make an Ansatz for the
  functional form of $C_0(\vec{r})$ that provides a good description of the data.
  For instance for the pion correlator that is found to be independent of the angles, a spherically symmetric
  Ansatz is used.
  We then perform a least squares fit to the lattice data of the sum given on the right hand 
  side of Eq.~(\ref{Eq:img_1st}) extracting the fit parameters of the function that
  describes $C_0(\vec{r})$. The corrected correlator 
  is then constructed by subtracting from the lattice data the images determined from the fitted function to
  obtain:
  \begin{equation}
    C^{\textrm{corr}}(\vec{r}) = C(\vec{r}) - \sum_{\left|\vec{n}\right|\in(0,\sqrt{3}]} C_0(\vec{r}+\vec{n}L).
\label{Eq:wf_corr}
  \end{equation}
  The Ans\"atze for $C_0(\vec{r})$ for each particle are summarized below:
  
  \begin{align}
    C_0^\pi &= A_0\exp{(-m_0r^\sigma)},\nonumber\\
    C_0^\rho &= \biggl[A_0\exp{(-m_0r^\sigma)}+A_1\exp{(-m_1r^\sigma})r^2 P_2(\cos\theta)\biggr]^2,\nonumber\\
    C_0^N  &= \textrm{same as for } \pi,\nonumber\\
    C_0^\Delta &= \textrm{same as for } \rho.
  \end{align}
  
  As can be seen, for the pion and the nucleon we take spherical functions.
  For the case of the $\rho$ 
  we have parametrized the correlator in such a way so that 
  an asymmetry, as seen in the uncorrected data, is allowed.
  For the $\Delta$, although no asymmetry can be seen within our statistical
  errors we use the same Ansatz as for the $\rho$ to see if the data allow
  for such a term. 
  
  Since the spatial part of the correlators is even 
  under reflection, only $L=0$ and $L=2$ angular momentum 
  quantum numbers are allowed. Thus for the $\rho$-meson and the $\Delta$
  we include an $L=2$ component by including the Legendre polynomial 
  $P_2(\cos{\theta})$. 
  In Table~\ref{Table:FitParams} we summarize the fit parameters obtained. The fact that for the spin projection
  $m_z=0$ $\rho$ state 
  the asymmetric term with coefficient $A_1$ is found non-zero and positive confirms 
  that the correlator is indeed 
  elongated along the $z$-axis (prolate) while the same parameter is consistently negative 
  for the $m_z=\pm1$ channels
  pointing to a correlator larger at the equator (oblate). 
  For the $\Delta$ the $A_1$ coefficient comes out negative for all quark masses
  albeit with a large statistical error not allowing any definite conclusions
  on the $\Delta$ shape.

  In Figs.~\ref{Fig:img_pn} and \ref{Fig:img_rd} we show a comparison between the raw lattice data and the
  lattice data after subtracting image contributions for the heaviest pion mass available. 
  As can be seen, the correction 
  procedure clearly compensates for the images, i.e. the spikes at $L/2$, $\sqrt{2}L/2$ and $\sqrt{3}L/2$ are 
  corrected for, leading to a smoother correlator that decreases more rapidly at the tails. 
  
  \bw
  
  \begin{table}[h]
    \caption{ $\langle x^2+y^2\rangle/2$, $\langle z^2\rangle$ and their difference for each particle at all three pion masses in fm$^2$, left for mesons and right for baryons. All errors are jack - knife errors.}
    \label{Table:Asymm}
    \begin{minipage}[t]{0.45\linewidth}
      \begin{tabular}[t]{cr@{.}lr@{.}lr@{.}l} 
        \hline
        \hline
        \multicolumn{1}{c}{$m_\pi^2$ (GeV$^2$)}&\multicolumn{2}{c}{$\langle x^2+y^2\rangle/2$} &\multicolumn{2}{c}{$\langle z^2\rangle$}&\multicolumn{2}{c}{$\langle z^2-(x^2+y^2)/2\rangle$}\\
        \hline
        \multicolumn{7}{c}{$\pi$}\\
        0.477 & 0&1449(6) & 0&1460(7) & 0&0011(8)\\
        0.259 & 0&1542(7) & 0&1531(9) &-0&0010(10)\\
        0.147 & 0&1529(7) & 0&1533(14)& 0&0005(18)\\
        \hline
        \multicolumn{7}{c}{$\rho$, $m_z=0$}\\
        0.477 &0&174(2) & 0&192(2) & 0&018(3)\\
        0.259 &0&188(4) & 0&196(6) & 0&007(7)\\
        0.147 &0&190(5) & 0&207(6) & 0&016(7)\\
        \hline
        \multicolumn{7}{c}{$\rho$, $m_z=\pm1$}\\
        0.477 & 0&183(1) & 0&173(2) & -0&009(2) \\
        0.259 & 0&199(2) & 0&186(2) & -0&013(2) \\
        0.147 & 0&200(4) & 0&193(5) & -0&007(6) \\
        \hline
        \hline 
      \end{tabular}
    \end{minipage}
    \begin{minipage}[t]{0.45\linewidth}
      \begin{tabular}[t]{cr@{.}lr@{.}lr@{.}l} 
        \hline
        \hline
        \multicolumn{1}{c}{$m_\pi^2$ (GeV$^2$)}&\multicolumn{2}{c}{$\langle x^2+y^2\rangle/2$} &\multicolumn{2}{c}{$\langle z^2\rangle$}&\multicolumn{2}{c}{$\langle z^2-(x^2+y^2)/2\rangle$}\\
        \hline
        \multicolumn{7}{c}{$N$}\\
        0.477 & 0&164(1) & 0&159(1) & -0&006(2) \\
        0.259 & 0&170(1) & 0&168(2) & -0&002(3) \\
        0.147 & 0&181(1) & 0&182(2) & 0&0008(31) \\
        \hline
        \multicolumn{7}{c}{$\Delta$, $m_z=\pm\frac{3}{2}$}\\
        0.477 & 0&177(1) & 0&172(1) & -0&005(2) \\
        0.259 & 0&182(1) & 0&180(2) & -0&001(2) \\
        0.147 & 0&195(2) & 0&198(3) & 0&003(4) \\
        \hline
        \hline 
      \end{tabular}
    \end{minipage}
  \end{table}

  \begin{figure}[h]
    \begin{minipage}{0.45\linewidth}
      \includegraphics[width = 1\linewidth]{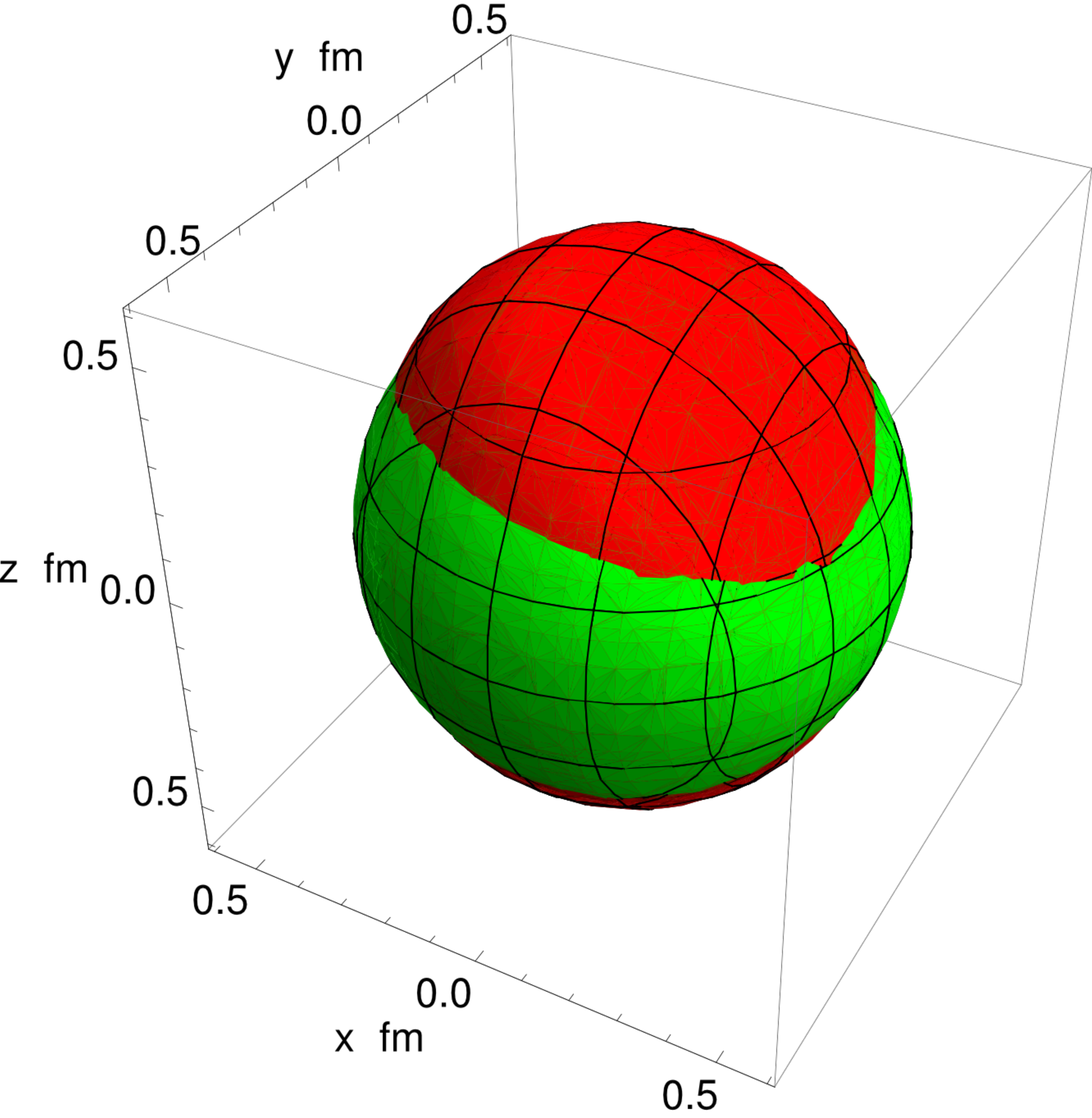}
      \caption{Three-dimensional contour plot of the $\rho$-meson, $m_z=0$ correlator (red or darker surface) compared to a sphere (green or lighter surface). The sphere radius is approximately 0.5~fm. The contour shows all $\vec{r}$ such that $C(\vec{r})=\frac{1}{2}C(0)$.}
      \label{Fig:3D_cont_rho0}
    \end{minipage}
    \hfill
    \begin{minipage}{0.45\linewidth}
      \includegraphics[width = 1\linewidth]{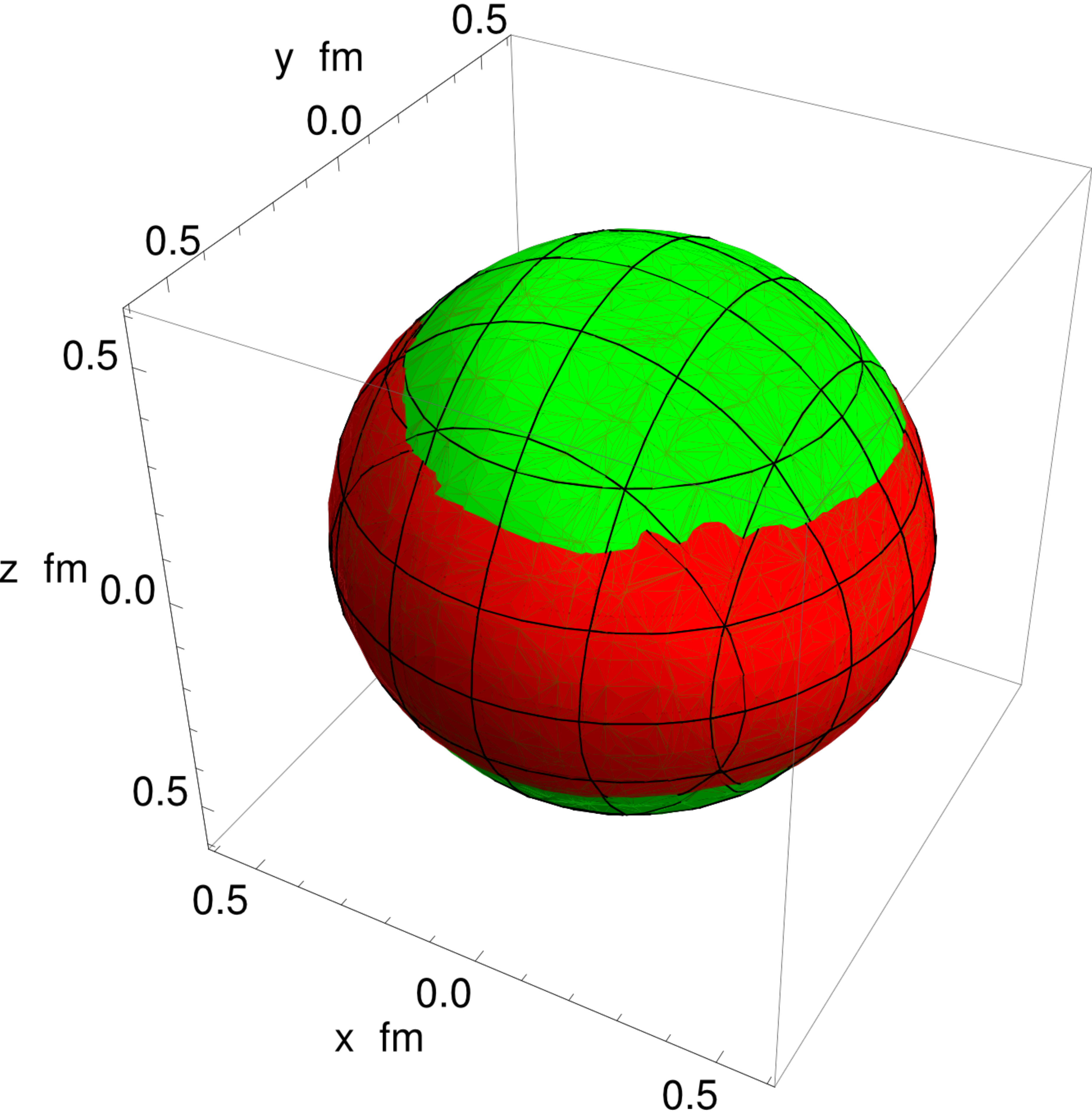}
      \caption{Three-dimensional contour plot of the $\rho$-meson, $m_z=\pm1$ correlator (red or darker surface) compared to a sphere (green or lighter surface). The sphere radius is approximately 0.5~fm. The contour shows all $\vec{r}$ such that $C(\vec{r})=\frac{1}{2}C(0)$.}
      \label{Fig:3D_cont_rho1}
    \end{minipage}
  \end{figure}
  
  \ew

  Having corrected the data for the nearest images we can now proceed to 
  a quantitative analysis of the particle charge distributions. In Table \ref{Table:Asymm} 
  we give $\langle x^2+y^2\rangle/2$, $\langle z^2\rangle $ and their difference for each particle 
  at each pion mass available.
  All errors are jack - knife errors. Here, the moments presented are computed using the corrected
  correlator:
  \be
  \langle\mathcal{O}\rangle = \frac{\sum_{\vec{r}}\mathcal{O}(\vec{r})C^{\textrm{corr}}(\vec{r})}{\sum_{\vec{r}}C^{\textrm{corr}}(\vec{r})}.
  \ee
  
  From the difference $\langle z^2-\frac{x^2+y^2}{2}\rangle$ we see once again that the $m_z=0$ projection of the 
  $\rho$ is larger along the $z$ axis while the $m_z=\pm1$ projections are larger along the equator. An additional observation here
  is that the asymmetry of the $m_z=\pm1$ projections is approximately half that of the $m_z=0$ projection, thus verifying the result 
  reached in Ref.\cite{Alexandrou:2002nn}. 
  For the case of the $\Delta$ on the other
  hand a spherical 
  distribution cannot be excluded, although for
  the two lightest pion masses we increase the statistics by 
  computing the correlators using the first and the second half temporal
  extent of the lattice and by using $N_r$=9 noise vectors
  for the smallest of the two values.
  
  The asymmetry in the $\rho$ is nicely represented
  by a three-dimensional contour plot.
  In Figs.~\ref{Fig:3D_cont_rho0} and \ref{Fig:3D_cont_rho1} we show 
  contour surfaces for the $\rho$-meson in the $m_z=0$ and $m_z=\pm1$ channels
  respectively, at the intermediate pion mass.
  The correlator is compared to a sphere centered at the origin. Once again we see that the $m_z=0$ state is 
  elongated along the poles while the $m_z=\pm1$ channels are flatter.

  \section{Summary and Conclusions}
  In this work we develop the formalism for the exact evaluation 
  of the equal time density-density correlators, which in the 
  non-relativistic limit reduce to the hadron charge distribution. The pion, $\rho$-meson, 
  nucleon and $\Delta$ density-density correlators are evaluated
  using dynamical Wilson fermions down to a pion mass of 384~MeV. 
  The all-to-all propagators
  needed 
  for the calculation of these correlators
  are computed using stochastic techniques combined with dilution. 
  Having the all-to-all propagators is required so that an explicit
  projection to zero momentum initial and final states is carried out. 
  In the meson-sector
  we implemented the one-end trick, which leads to a significant
  improvement in the accuracy with which the density-density
  correlators are obtained. 
  This improved accuracy is needed to
  conclude with certainty that the 
  the $\rho$-meson is deformed. The $\rho$ is found to be a prolate
  when in the spin projection zero state and an oblate
  in the spin projection $\pm 1$ state. 
  This result corroborates previous studies 
  where the density-density correlator of the
  $\rho$ was calculated without explicit zero-momentum projection and with less accuracy~\cite{Alexandrou:2002nn}.
  It is also in agreement with a negative quadrupole form factor calculated recently on the lattice~\cite{Hedditch:2007ex}.
  For the baryons a spherical distribution can not be excluded
  given the present statistical errors despite increase in statistics.
  
  Finite spatial volume effects affect mainly
  the long distance behavior of the correlators.
  By adopting an Ansatz for the asymptotic dependence 
  of the correlators we correct for these finite volume effects by subtracting
  the first image contributions. The functional form determined from fits
  to the corrected data confirm a deformed shape for the $\rho$ meson.
  For the $\Delta$, although the fits allow for a small deformation, 
  the statistical error is too large to exclude a spherical distribution.
  Further improvements in the evaluation of all-to-all propagators such
  as combination of stochastic techniques and lower eigenmode projection
  are currently being investigated by a number of groups with
  promising results~\cite{Morningstar:2008} that have potential
  application in the study of baryon density-density correlators.
  
  \begin{acknowledgments}
    G. K. would like to acknowledge support by the 
    Cyprus Research Promotion Foundation. The computations 
    for this work were partly carried out on the IBM Power6 575 
    machine at NIC, Julich, Germany.
  \end{acknowledgments}
  
  \bibliography{Bibliography}
\end{document}